\documentclass{aa}  
\usepackage{natbib}
\usepackage{xcolor}
\usepackage{graphicx}
\usepackage{natbib}
\bibpunct{(}{)}{;}{a}{}{,}
\usepackage{txfonts}
\usepackage{soul}
\newcommand{\blos}{B$_{\parallel}$}

\newcommand{\divbo}{$\nabla \cdot \mathbf{B} =0$}

\newcommand{\bperp}{B$_{\perp}$}
\newcommand{\us}{TPBK18}
\newcommand{\KS}{$\chi ^2$}
\usepackage[utf8]{inputenc}
\usepackage{fourier} 
\usepackage{array}
\usepackage{makecell}

\begin{document}

\title{Bow Magnetic Morphology Surrounding Filamentary Molecular Clouds?}
 \subtitle{3D Magnetic Field Structure of Orion-A}
   \author{M. Tahani\inst{1,2},
R. Plume\inst{2},
J. C. Brown
          \inst{2}
          \and
          J. D. Soler \inst{3}
          \and
          J. Kainulainen \inst{3,4}
          }
      
   \institute{Dominion Radio Astrophysical Observatory, Herzberg Astronomy and Astrophysics Research Centre, National Research Council Canada, P. O. Box 248, Penticton, BC V2A 6J9 Canada\\
   \email{mehrnoosh.tahani@nrc.ca}
   \and Physics \& Astronomy, University of Calgary, Calgary, Alberta, Canada\\
   \email{mtahani@ucalgary.ca}
    \and Max-Planck-Institute for Astronomy, K\"{o}nigstuhl 17, 69117 Heidelberg, Germany
    \and Chalmers University of Technology, Department of Space, Earth and Environment, SE-412 93 Gothenburg, Sweden}

   \date{Received ; accepted }
  
\titlerunning{Bow Magnetic Field Morphology?}
\authorrunning{M. Tahani et al.}

\abstract
{Using a method based on Faraday rotation measurements, \cite{Tahanietal2018} find the line-of-sight component of magnetic fields in Orion-A and show that their direction changes from the eastern side of this filamentary structure to its western side. Three possible magnetic field morphologies that can explain this reversal across the Orion-A region are toroidal, helical, and bow-shaped morphologies.}{In this paper we construct simple models to represent these three morphologies and compare them with the available observational data to find the most probable morphology(ies).}{To compare the observations with the models, we use probability values and a Monte-Carlo analysis to determine the most likely magnetic field morphology among these three morphologies.}{We find that the bow morphology has the highest probability values and that our Monte-Carlo analysis suggests that the bow morphology is more likely.}{
We suggest that the bow morphology is the most likely and the most natural of the three morphologies that could explain a magnetic field reversal across the Orion-A filamentary structure (i.e., bow, helical and toroidal morphologies).}{}

\keywords{magnetic fields --- ISM: clouds --- ISM: magnetic fields ---  stars: formation} 

\maketitle

\section{Introduction}
\label{Intro}

Theoretical studies suggest that the orientation and strength of magnetic fields are dynamically important in the formation and evolution of filaments and filamentary Molecular Clouds \citep[MC; e.g.,][]{Hartmannetal2001,McKeeOstriker2007,  HennebelleFalgarone2012, SeifriedWalch2015, HennebelleInutsuka2019}. Observations of magnetic fields indicate that there is a coupling between the matter and the magnetic fields in these regions ~\citep[e.g.,][]{Crutcher2012, Lietal2013, PlanckXXXV}. These studies are not conclusive, however, and we require more studies, techniques, and observations to constrain the role and effects of magnetic fields in these star forming regions. 

To explore magnetic field morphologies, a number of observations have been done exploiting dust polarization and Zeeman measurements. For example, the Planck Collaboration has observed the plane-of-sky component of the magnetic field (\bperp) in different MCs \citep[][hereafter PXXXV]{PlanckXXXV}. Zeeman measurements have been successful in obtaining the magnetic field component parallel to the line-of-site (\blos) in star forming regions \citep[e.g.,][]{Goodman1989, Crutcher1999APJ520, Crutcher2005}. \cite{TrolandCrutcher2008} report the largest number of OH Zeeman observations toward MCs. A recent study by \citet[][hereafter \us]{Tahanietal2018} propose a new method based on Faraday rotation to determine \blos\ in MCs. Prior to TPBK18 method, however, Faraday rotation studies were traditionally associated with the large scale Galactic magnetic field and mostly ionized medium.

Each of these observations only provides one component of the magnetic fields, i.e., \blos\ or \bperp. Probing the three dimensional field morphology requires combining the individual measurements (Zeeman splitting, Faraday rotation, and dust polarization technique), developing a new technique, or using geometrical models. Recently \cite{Chenetal2018} proposed obtaining the three-dimensional (3D) magnetic field morphology using dust polarization observations, based on the statistical properties of the observed polarization fraction. Another technique to obtain the 3D magnetic field in star-forming regions involves 
combining dust polarization data, Zeeman measurements, and ion-to-neutral molecular line width ratio measurements \citep[e.g.][]{Houdeetal2002}. However, due to a degeneracy between the inclination angle and the strength of \bperp\ and ambiguity in the direction of \bperp, \cite{LiHoude2008} and \cite{Houde2011} suggested that this technique cannot be widely used. Geometrical models have also been used to provide insights to the 3D magnetic field morphologies of filamentary structures \cite[e.g.][]{PlanckXXXII2016, PlanckXXXIII2016}. 
The method of \us\ holds the promise of wider exploration of \blos\ with future RM surveys with more sensitive measurements and reduced uncertainties. This will improve our understanding of the 3D magnetic field morphologies associated with star forming MCs.

\us\ find \blos\ in  Orion-A, Orion-B, California, and Perseus MCs.  They find that,  within the uncertainties, their obtained values are consistent with the available molecular Zeeman measurements in these regions. They reveal that in California and Orion-A MCs, the direction of line-of-sight (LOS) magnetic fields reverses from one side of these filamentary structures to the other side. This magnetic field reversal in Orion-A has been previously observed \citep{Heiles1997, PressRelease} and has been interpreted in different ways.

Three magnetic field morphologies that can explain this direction-change of \blos\ across filamentary structures are toroidal, helical, and bow morphologies. The {\em helical/toroidal} morphology has been investigated in a number of theoretical studies \citep[e.g.,][]{ShibataMatsumoto1991, Nakamuraetal1993, Hanawaetal1993, Matsumoto1994, FiegePudritzI2000,FiegePudritzII2000, SchleicherStutz2017,  Reissletal2018}. \cite{FiegePudritzI2000} and \cite{FiegePudritzII2000} study the fragmentation length-scale, stability, density profile, and mass per length of filamentary MCs and, based on observational constraints, they suggest that many filamentary structures are likely wrapped by helical magnetic fields.
Other studies of Orion-A, indirectly suggest a helical magnetic field morphology for this region \citep[e.g.,][]{JohnstonBally1999, Matthews2001, Buckleetal2012, Contrerasetal2013, StutzGould2016, Hoqetal2017}. For example, by using the Virial mass per length obtained by \cite{FiegePudritzI2000} for a cylindrical filament threaded by a helical magnetic field, \cite{Buckleetal2012} show that the integral shaped filament in Orion-A is too massive for thermal or turbulent support. Therefore, they suggest that the mass and morphology of the integral shaped filament (a small region within our defined Orion-A) is consistent with a Virial model of a filamentary structure threaded by a helical magnetic field.

\begin{figure}
\centering
\includegraphics[scale=0.45, trim={3.5cm 4.5cm 4.5cm 4cm},clip]{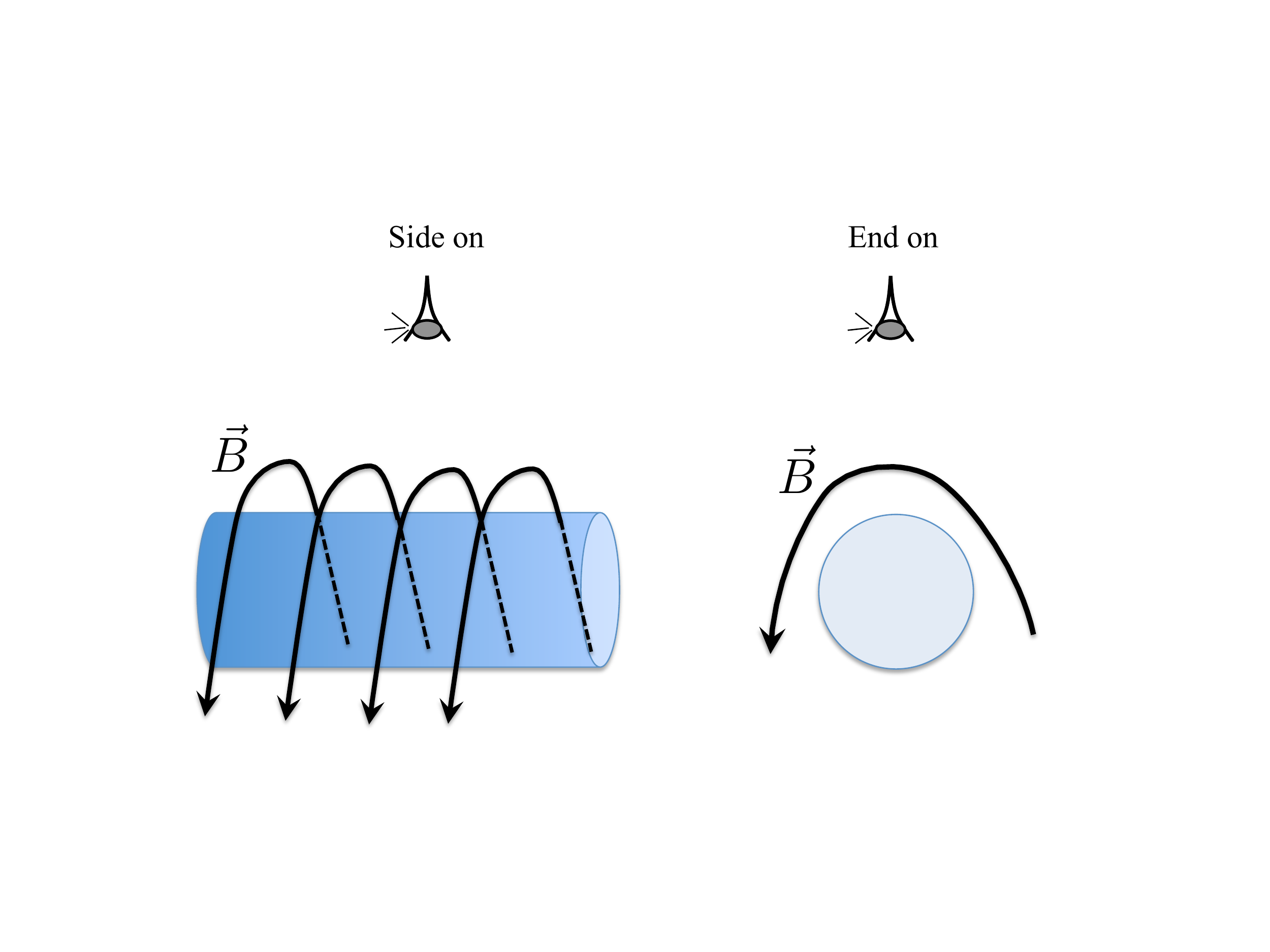}
\caption[Wrapping of magnetic fields lines around a cylindrical filamentary structure]{Wrapping of magnetic fields lines around a cylindrical filamentary structure (blue circle), as seen by the observer.  This concept could provide an explanation for the observations seen in Orion-A, interacting with the Orion-Eridanus superbubble. The magnetic field line is represented as the curved black line.}
\label{Wrapped}
\end{figure}

\cite{Heiles1997} proposes an alternative explanation, which associates the magnetic reversal in Orion-A to the Eridanus shock and its interaction with the dense MC. In this mechanism, the magnetic fields bend around the Orion MC, as illustrated in Fig.~\ref{Wrapped}. The ambient Galactic \blos\ in this region is, in general, towards the observer and interacts with the superbubble, generating the reversal \citep{Heiles1997}. \cite{Heiles1997} suggests that a reversal in this region can be observed even without the presence of a MC, however, the existence of a MC in this region makes this reversal sharper. This model is supported by the Planck observations considered in \cite{Soleretal2018}, who map \bperp\ lines in the Orion-Eridanus super-bubble and suggest that due to the large-scale shape of the magnetic field lines, \bperp\ interacts with, and is influenced by, the Orion-Eridanus superbubble.

We refer to this morphology, regardless of how it is formed, as a {\it bow} morphology. This general morphology has been recently investigated in some theoretical studies \citep{Gomezetal2018,Inoueetal2018,Reissletal2018}, as well as observational studies \citep{Liuetal2018}, and has been also referred to as a U-shape or a  pinched field.

Understanding the overall 3D magnetic field morphology, and consequently identifying which of these proposed magnetic configurations are responsible for the observed magnetic field reversals, can potentially help us answer questions pertaining to the role of magnetic fields in star formation. For example, helical fields can allow for more mass accumulation in the filamentary structure by stabilizing the cloud against self gravity \citep{Buckleetal2012}. Additionally, since the magnetic fields must form closed loops (zero-divergence), determining the 3D magnetic field morphology in and around these filaments can provide information regarding the influence that the surrounding environment has on the formation and evolution of filamentary structure. 

In this paper we investigate the large-scale 3D magnetic field in Orion-A (the entire southern complex). For this purpose we use PXXXV dust polarization data and the \us\ \blos\ results. We discuss each of these datasets in Sec.~\ref{data} and describe our methodology to couple these two magnetic fields results in Sec.~\ref{method}.  Finally, we discuss our results and interpretation. 
The goals of \us\ and this paper  are to provide techniques to obtain the large-scale structure of magnetic fields around star forming MCs. The broad-perspective intent is to show what can be done with the existing data, as well as to set the stage for future, higher sensitivity RM catalogs, which will be obtained by next generation surveys and telescopes.

\section{Observations}
\label{data}
Below we discuss the data we use for the line-of-sight and plane-of-sky components of the magnetic field in our study.

\subsection{Line-of-sight magnetic field component}

For the parallel component of the magnetic field (\blos), we use the directions and magnitudes found in \us. They calculate \blos\ using \cite{Tayloretal2009} rotation measure (RM) catalog, in combination with an electron column density estimated for each RM source from a chemical evolution code and \cite{Kainulainenetal2009} extinction maps. Additionally, they find \blos\ reversal across two regions of California and Orion-A molecular clouds. \us\ explains in detail how to find the \blos\ strengths and their uncertainty values. These uncertainties are calculated for each data point separately, and are due to the uncertainty propagation of all the required variables (input parameters) in determination of \blos\ strength.

\begin{figure}
\centering
\includegraphics[scale=0.6, trim={5cm 0.5cm 1.5cm 1cm},clip]{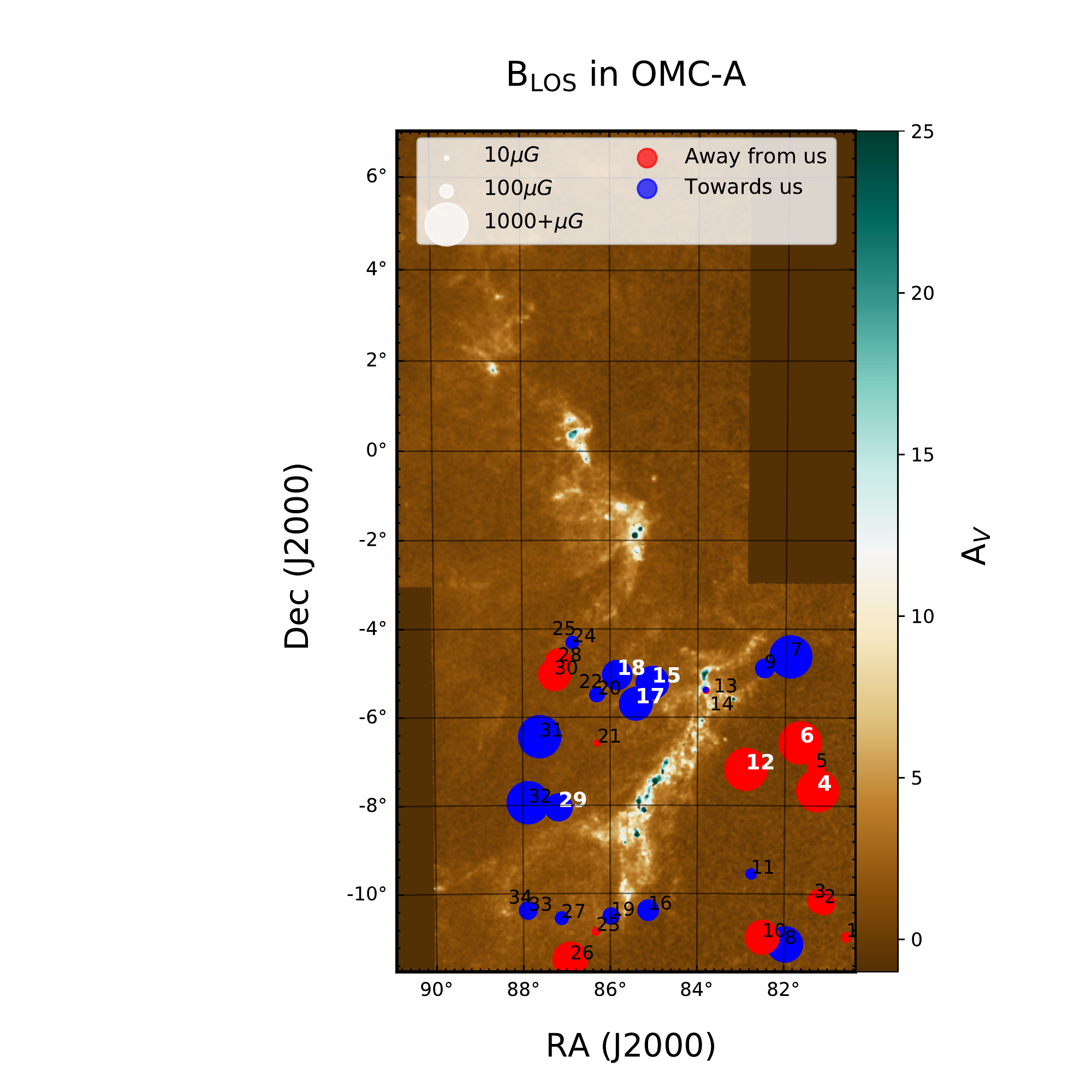}
\caption[\blos\ in Orion-A]{\blos\ in Orion-A. Blue (red) circles show magnetic fields toward (away from) us. The size of the circles indicate the magnitude of magnetic field. Sources numbered in bold white indicate the sources used for this study. 
Points 3, 4, 6, 7, 8, 12, 15, 17, 18, 19, 29, 31, 32, and 34 in this figure have errors less than 100\%. Points 4, 6, 12, 15, 17, 18, and 29 are considered in this study.
}
\label{OrionBlos}
\end{figure}

In this paper, we consider only the magnetic field around Orion-A and use the data points that provide a solid direction for \blos\ (either towards or away from us). Consequently, if a \blos\ value has an uncertainty bigger than the value of \blos\ itself, we do not consider that data point in our analysis. That is, we use the B values with uncertainties $< $ 100\% so that the direction of the field is well determined. For example, point 13 with $\text{B}_{\parallel} = -23 \pm 38\,\mu$G (an uncertainty $>$  100\%) changes direction within its uncertainty range. We do not consider this point in our analysis, since its direction is highly uncertain. Fig.~\ref{OrionBlos} shows all the data points determined in \us. Points 3, 4, 6, 7, 8, 12, 15, 17, 18, 19, 29, 31, 32, and 34 in this figure have errors less than 100\%. Points 4, 6, 12, 15, 17, 18, and 29, are considered in this study, since these are relatively close to the filament (filamentary structure) and are within the main axis length of the filament (see Sec. \ref{integrate}). In other words, these are the points with nearby \bperp .

\subsection{Plane-of-sky magnetic field component}
\label{PlanckData}
For the plane-of-sky component (\bperp), we use the PXXXV determination of the magnetic field orientation as projected on the plane of the sky. As explained in PXXXV, we find the magnetic field lines from the thermal emission of interstellar dust observed by Planck\footnote{\url{https://www.esa.int/Our_Activities/Space_Science/Planck}} at 353 GHz taken at an original resolution of $4'.8$ and then smoothed with a $15'$ FWHM Gaussian beam to guarantee signal-to-noise ratio of at least 3.

To find the average \bperp\ value in the Orion-A region we use the Planck dust polarization data along with the method of Davis-Chandrasekhar-Fermi (DCF) plus the structure function (DCF+SF) described by \cite{Hildebrandetal2009}. 
To obtain the DCF components, we use the Planck 353 GHz data for the angle dispersion and CO observations for the density and line-of-sight velocity dispersion. The value we obtain is $180 \pm 90 \mu$G. This is a refined value tailored for Orion-A only, compared to what is mentioned in Table D.1 in PXXXV, which is for the entire Orion complex. Similar to \blos, the \bperp\ uncertainty is as the result of uncertainty propagation of the input parameters (density and velocity dispersion) into determination of \bperp. Even though these values do not represent \bperp\ magnitude for each data point of \us\ Orion-A map, where there is CO emission, the \bperp\ \emph{orientation} at each point is determined from the Stokes parameters at that point. 

\section{Methodology}
\label{method}

To investigate the validity of possible magnetic field morphologies, we model the different proposed configurations and compare the results of averaged magnetic fields along the LOS to those of observations. 

\subsection{Integrating the observations with models}
\label{integrate}

In order to compare the modeled values with the observations, we need to find a \bperp\ value and orientation that can be associated with the \blos\ at each point from \us. We do this by finding the closest PXXXV \bperp\ to the \blos\ points. If no \bperp\ value exists within $15'$ (the size of the smoothed Planck polarization beam) of the \blos\ point, we discard that \blos\ point from our analysis. This also means that the \blos\ data point is too far from the filament (points 3, 8, 19 in Fig.~\ref{OrionBlos}). Points 31, 32, 34 and 7 are also too far away from the filament axis. Therefore, we carry out this study with points numbered 4, 6, 12, 15, 17, 18, and 29, as indicated by the white numbers in Fig.~\ref{OrionBlos}. 

After finding the closest \bperp\ point to our \blos\ point, we find the orientation of \bperp\ by determining its polarization angle using:
\begin{equation}
\psi = \frac{1}{2} \arctan(-U, Q),
\end{equation}
where $\psi$ is the polarization angle in the IAU convention, and U and Q are the Stokes parameters obtained from Planck observations of dust thermal radiation.
We add a $\pi/2$ to $\psi$ to account for the fact that the magnetic field orientation ($\phi$) is orthogonal to the polarization angle. These angles are initially found with respect to the Galactic coordinate. To analyze the 3D geometry of the magnetic field with respect to the filament itself, we convert (rotate) these angles to the frame of reference of the filament itself. 

\begin{figure}
\centering
\includegraphics[scale=0.7, trim={0cm 0cm 0cm 0cm},clip]{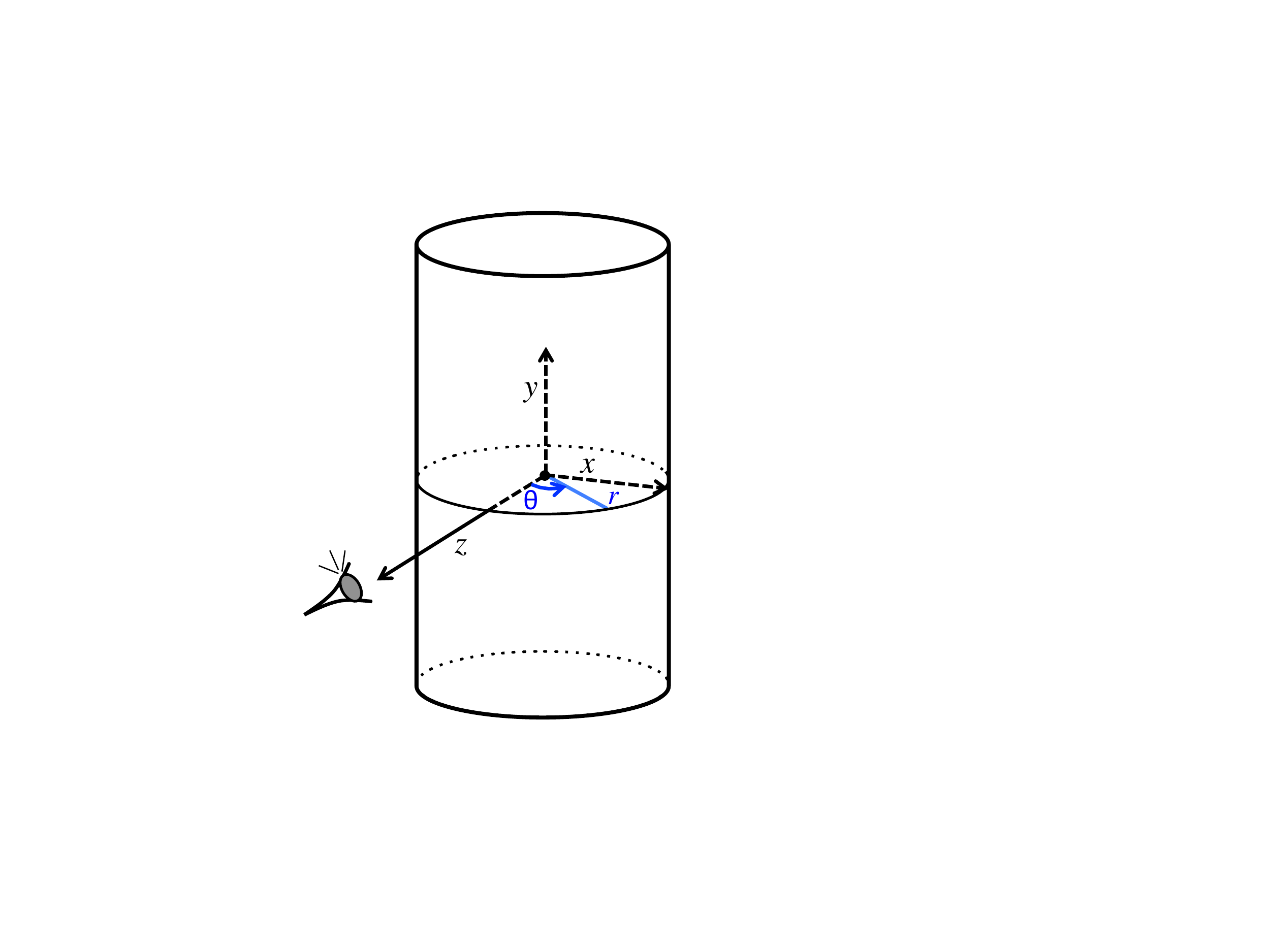}
\caption[Geometry and coordinates for a filament]{Geometry and coordinates for a filament. The x-axis is radial to the filament, and in the plane of the sky, y-axis is parallel to the long axis of the filament, and z-axis is radial, towards the observer.}
\label{Cylindre}
\end{figure}

We set the filament frame of reference such that the x-axis is radial to the filament and in the plane of the sky (i.e., the short axis of the filament in the plane of the sky from east to west), y-axis is parallel to the long axis of the filament in the plane of the sky, and z-axis is radial, towards the observer, as illustrated in Fig.~\ref{Cylindre}. We can then write the magnetic field vector at each point in this frame of reference as ${\bf B} = (B_{x}, B_{y}, B_z)$. Compared to the Galactic system, this new frame of reference is rotated by $\xi$, where $\xi$ is the angle the filament's minor axis makes with decreasing Galactic longitude axis, i.e., the orientation of the x-axis of the filament with respect to East-West direction of the map measured in a clockwise direction.

In this system, $|B_x| = |B_{Planck}\sin(\xi+ \phi)|$ and $|B_y| = |B_{Planck}\cos(\xi+ \phi)|$, where $\phi$ is the magnetic angle in the plane of the sky and $B_{Planck} = 180 \pm 90\, \mu$G. The equations are cast as absolute values because we do not yet have any information regarding the positive/negative signs of $B_x$ and $B_y$ that provide the direction of \bperp . While the {\em orientation} of \bperp can be identified from the Stokes parameters, the actual {\em direction} (+ or $-$ along that orientation) cannot be determined.  $B_z =$ \blos\ and does include the positive and negative signs of \blos . 

Fig.~\ref{ObservedBRatios} shows the resultant magnetic fields in the Cartesian frame illustrated in Fig.~\ref{Cylindre}.  In this figure, we take the ratio of the observed magnetic field components (i.e., $B_x$, $B_y$, $B_z$) to the total estimated observed magnetic field strength ($B_{total} = \sqrt{B_x^2 + B_y^2 + B_z^2}$). Considering the ratios helps us reduce the number of free parameters in the models. To distinguish between morphologies, we compare these observed magnetic field strengths to those predicted by  our  three models as described in sections below.

\subsection{Modeling a toroidal and a helical 3D magnetic field}
\label{HelicalModel}

We model a filament with a simple toroidal or a helical magnetic field using the following set of equations:

\begin{equation}
\begin{aligned}
B_x = & \frac{B_0 \text{R} \cos(\theta)}{r^{\alpha}}= \frac{B_0 \text{R}z}{(x^2 + z^2)^{\frac{\alpha +1}{2}}}, \\ 
B_y = &\frac{B_1\text{R}}{r^{\alpha}}= \frac{B_1\text{R}}{(x^2 + z^2)^{\frac{\alpha}{2}}},\\ 
B_z = & \frac{-B_0 \text{R}\sin(\theta)}{r^{\alpha}} = \frac{-B_0 \text{R}x}{(x^2 + z^2)^{\frac{\alpha +1}{2}}},
\label{HelicalBEq}
\end{aligned}
\end{equation}
where the x, y, and z directions follow the convention set in Sec. \ref{integrate} and Fig.~\ref{Cylindre}.
R is a scaling factor with units of [length]$^{\alpha}$ and $r=\sqrt{x^2+z^2}$ - the radial distance from the filament mid-plane in the X-Z plane (i.e., perpendicular to the long axis). $B_0$ and $B_1$ carry information about the strength of the toroidal and poloidal components of magnetic field respectively. The parameter $\alpha$ reflects the rate at which the  magnetic field strength decreases with  distance from the filament.

To have a purely toroidal magnetic field, or a helix with a small pitch angle,  $B_1$ needs to be either zero or very small compared to $B_0$. In these equations, $\frac{x}{(x^2 + z^2)^{1/2}}$ and $\frac{z}{(x^2 + z^2)^{1/2}}$ replace $\sin\theta$ and $\cos\theta$. 

Since a helical field threading a filament must be caused/produced by the filament itself, we assume its strength drops with distance from the filament's main axis. The scaling factor R, therefore, arises from the relation between the strength of the magnetic field and its distance from the filament.  This ensures that the units of both sides of these equations are identical (i.e., units of magnetic field strength).

To understand the rationale behind the magnetic strength decreasing with 1/r$^{\alpha}$, we consider the following: the relation between the magnetic field and MC gas density; the observed density profile for the Orion-A filament; the necessity to keep the magnetic field divergence free.

The relationship between the magnetic field strength and the gas density has been investigated both theoretically and observationally. Theoretically, \cite{Tritsisetal2015} find that   $B(\rho) \propto \rho^{1/2}$ is preferred, where $B(\rho)$ is the magnetic field as a function of the mass volume density and $\rho$ is the mass volume density. Observationally, \cite{Crutcher2010} propose a ratio of $B(\rho) \propto \rho^{2/3}$  when the magnetic energy is small compared to the effects of gravity. In our models, we consider both $B(\rho) \propto \rho^{1/2}$ and $B(\rho) \propto \rho^{2/3}$. 

The density profile of the Orion-A filament has also been studied both observationally and theoretically and is proposed to have a Plummer-like form \citep[e.g.][]{Saljietal2015}. The standard Plummer profile is described by:
\begin{equation}
\rho (r) = \frac{\rho_c}{\big[1 + (r/r_{flat})^2\big]^{p/2}},
\label{PlummerDensity}
\end{equation}
where $\rho_c$ is the central density of the filament, $r$ is the distance (radius) from the filament axis, and $r_{flat}$ is the characteristic radius defining the central region, where the density profile flattens. The exponent  $p$ is observationally determined and sets the density drop-off rate. 
However, we use the more recently defined density profile suggested by \citet[See their equation 5]{StutzGould2016}, where they suggest that $\rho(r) \propto r^{-13/8}$.  This density profile not only fits the observations, it  also keeps our magnetic   field divergence-free (i.e.,  $ \frac{\partial B_x}{\partial x} + \frac{\partial B_y}{\partial y} + \frac{\partial B_z}{\partial z}= 0$ in Equation \ref{HelicalBEq}).  We extend this density profile to the larger Orion-A filamentary structure.
Combining B$(\rho)$ with $\rho(r)$, we obtain a relation for B$(r)$ in which  the magnetic field strength decreases 
as 1/r$^{\alpha}$, where $\alpha$ can be either 13/16 or 13/12 depending on whether we use the \cite{Tritsisetal2015} or the \cite{Crutcher2010} relations.

\subsection{Modeling a bow magnetic field morphology}
\label{WaveModel}

The simplest divergence-free magnetic field relation one can use to describe a bow morphology is:
\begin{equation}
\begin{aligned}
B_x = & B_1, \\
B_y = & B_2,\\
B_z = & B_0\frac{x}{\text{R}^{\prime}}, 
\label{WrappedB}
\end{aligned}
\end{equation}
where $B_0$, $B_1$, and $B_2$ are free parameters with positive values which provide the strength of magnetic field. R$^{\prime}$ is a free parameter with the unit of length to ensure that the relation for the z-component of magnetic field in equation~\ref{WrappedB} at both sides have the same units (see Sec.~\ref{BowAnalysis} for a range of values explored for these free parameters). 
$B_x$ and $B_y$ remain constant and in the direction of the ambient magnetic field, whereas $B_z$ varies with x.  Since the value of the x-coordinate varies from positive to negative across the filament, this implies that $B_z$ reverses direction from one side of the filament to the other.

It is important to note that these equations represent single magnetic field lines. If we follow one field line as it wraps around a filament, we will see that the z-component (towards the observer) of the field is zero directly in front of, and behind, the filament.  As we increase the radial distance from the filament (x-direction), the z-component will become larger. Thus, this relation of $B_z$ with $x$ represents the wrapping of the field lines around the filament, as well as the reversal.

Based on the model proposed by \cite{Heiles1997}, we assume that in the bow morphology the magnetic field components  $(B_{x}, B_{y}, B_z)$ represent field lines that do not penetrate the dense filament but, instead, represent the ambient magnetic field. Since they are ambient field lines, we do not expect a change of B-field strength as a function of distance from the filament's long-axis. Therefore, we make the following assumptions: in the above set of equations for the bow model, the overall magnetic field strength does not decrease with distance; $B_1$ = $B_2$  represents the ambient magnetic field. Thus, we only explore different ratios of $B_0$ and $B_1$ (see Sec.~\ref{discussion}). 

If we forgo the \divbo\ requirement, we can use the equations provided by \cite{Reissletal2018}, which model the bow morphology in a slightly different manner:
\begin{equation}
\begin{aligned}
B_x = & \frac{B_0(x,z)}{1+U^2(x,z)}, \\
B_y = & ~ 0 , \\
B_z = & \frac{B_0(x,z)U(x,z)}{1+U^2(x,z)}, \\
\label{Reissl}
\end{aligned}
\end{equation}
with
\begin{equation}
\begin{aligned}
& U(x,z) =  -5x(2-z)^2~e^{-8x^2}, \\
& B_0(x,z) =  \frac{B_0}{(1+(r/r_{flat})^2)^{-0.6\beta}};  & r = \sqrt{x^2+z^2}.
\label{ReissleInfo}
\end{aligned}
\end{equation}
$r_{flat}$ is the characteristic radius of the Plummer-like density profile (i.e., where the profile becomes flat close to the center of the filament). The parameter $\beta$ controls the slope of the density drop-off in the outer regions. \cite{Reissletal2018} use $\beta$ = 1.6, which is the average value of $\beta $ from the \cite{Arzoumanian2011} density profile. $B_0$ controls the total strength of the magnetic field.  

\subsection{Comparing observed 3D magnetic field strengths with models}
\label{ModelComparison}

Since, along any given LOS we see a superposition of many different field lines, the observed magnetic field components  $(B_{x}, B_{y}, B_z)$ are averages over the LOS.  To account for this in our models, we  find the average magnetic field components along the LOS for each x, y, and z component of the field using the following integral:
\begin{equation}
<B_i>(x/\text{L}) = \frac{\int^{+\frac{\text{L}}{2}} _{-\frac{\text{L}}{2}} B_i(x,z) dz}{\text{L}},
\label{average}
\end{equation}
where L is the integration length along the LOS and $i$ indicates $x$, $y$, or $z$ components.  
Therefore, $<B_i>(x/L)$ is the average value of the $x$, $y$, or $z$ component of the magnetic field at a radial distance of $x$ from the filament long axis (the y-axis), and averaged over the line-of-sight distance of L. 
$B_i(x,z)$ is the magnetic field components from each models in equations \ref{HelicalBEq}, \ref{WrappedB}, and \ref{Reissl}. This calculation is performed  for each of the models (toroidal/helical/bow) separately.
Thus, equation~\ref{average} provides $<B_x>$, $<B_y>$, and $<B_z>$ as a function of $x$ (perpendicular distance to the filament axis) and L (integration path). We explore these averages as a function of $x$/L to avoid the need to convert $x$ and L to real distances and reduce the number of free parameters in the analysis.
We set this integration distance (L) to be half of the distance of the point~17 from the filament axis, because we assume this distance represents the MC's thickness along the LOS. 
One should note that finding the observed \blos\ values do not require the actual integration distance, since \us\ use the column density values in instead of the volume densities. Information about the line-of-sight integration distance is already embedded in the column densities. 

To compare the models with the observed values, we determine the perpendicular distance (x) of each of our selected points in Fig.~\ref{OrionBlos} from the filament axis. Additionally, we find the ratio of each x, y, and z component to the total (averaged along the LOS) magnetic field - both in the models and in the data, i.e., $(\frac{<B_{x}>}{<B_{total}>}, \frac{<B_{y}>}{<B_{total>}}, \frac{<B_{z}>}{<B_{total}>})$, where $B_{total} = \sqrt{<B_x>^2 + <B_y>^2 + <B_z>^2}$. These represent the ratios that should be observed ideally without experimental uncertainties and systematic observing biases.
We then compare the above-mentioned ratios between the observed and modeled magnetic field components in order to determine which model is most consistent with the observed data.  We perform this analysis in Sec.~\ref{results}. 

\begin{figure}
\centering
\includegraphics[scale=0.42, trim={0cm 0cm 1cm 1.5cm},clip]{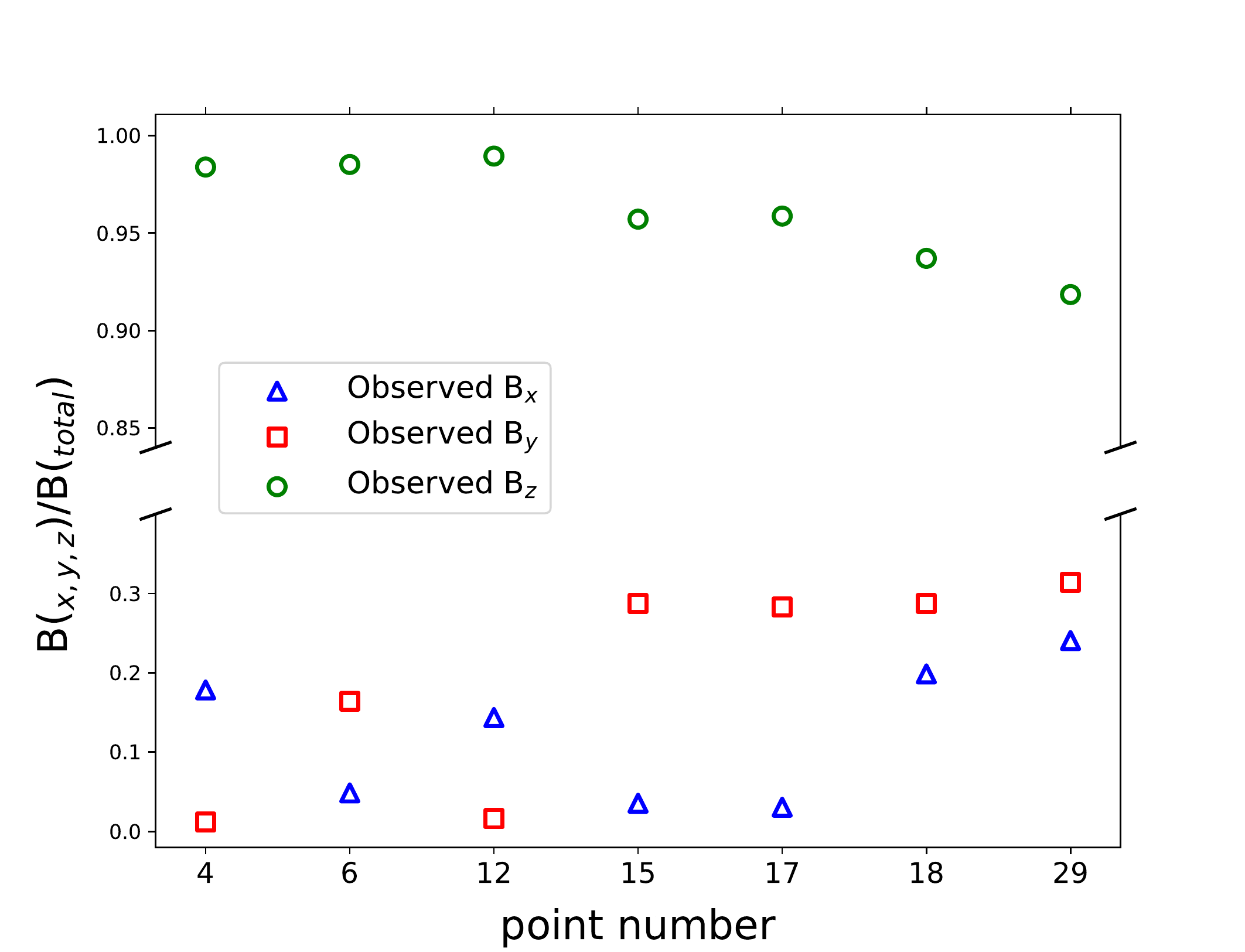}
\caption[Observed $B_x$, $B_y$, and $B_z$]{Observed $B_x$, $B_y$, and $B_z$ using the frame system depicted in Fig.~\ref{Cylindre} as well as \bperp\ and \blos\ values obtained by PXXXV and \us\, respectively. The y-axis shows the ratio of each component of magnetic field to the total value and the x-axis refers to the numbered data points labeled in Fig.~\ref{OrionBlos}.}
\label{ObservedBRatios}
\end{figure}

To compare the data with the models we employ a  \KS\ approach, using the following relation for each component of the magnetic field.

\begin{equation}
\begin{aligned}
&\chi ^2 _i=  \sum_{j=1}^{N} \frac{\bigg(\text{observed } \frac{<B_i>}{<B_{tot}>} - \text{modeled} \frac{<B_i>}{{<B_{tot}>}}\bigg)^2_j}{\delta ^2_j},\\
&\chi ^2 _{tot} = \chi ^2 _x + \chi ^2 _y + \chi ^2 _z,
\end{aligned}
\label{KS}
\end{equation}
where $i$ represents the $x$, $y$, or $z$ component of the magnetic field. N is the number of observational points in the study (in this paper $N= 7$), and $\chi ^2 _{tot}$ is the total $\chi ^2$.  ``Observed~$\frac{<B_i>}{<B_{tot}>}$'' is the observed ratio of each component of magnetic field to the total magnetic field as plotted in Fig.~\ref{ObservedBRatios}.  ``Modeled~$\frac{<B_i>}{<B_{tot}>}$'' is the LOS average of each model as explained in Sec.~\ref{ModelComparison}. $\delta$ represents the uncertainties of the magnetic field strengths determined in \blos\ and \bperp\ combined together. This $\delta$ is the unit-less relational uncertainty for the ratios as follows:
 \begin{equation}
 \delta = \sqrt{(\frac{\delta B_{||}}{B_{||}})^2 + (\frac{\delta B_{\perp}}{B_{\perp}})^2},
 \end{equation}
where $\delta B_{||}$ is the uncertainty of \blos\ and $\delta B_{\perp}$ is the uncertainty of \bperp. For example, for point number 29 in Fig.~\ref{OrionBlos}, with $\text{B}_{\parallel} = -418 \pm 308\,\mu$G and $\text{B}_{\perp} = 180 \pm 90\,\mu$G, $\delta$ is calculated as follows:
 \begin{equation}
 \delta = \sqrt{(\frac{308}{418})^2 + (\frac{90}{180})^2} = 0.89.
 \end{equation}

\section{Results of modeling}
\label{results}

Fig.~\ref{ObservedBRatios} demonstrates the ratio of the observed magnetic field components (i.e., $B_x$, $B_y$, $B_z$) to the total estimated observed magnetic field strength ($B_{total} = \sqrt{B_x^2 + B_y^2 + B_z^2}$), in the filament's Cartesian frame shown in Fig.~\ref{Cylindre}. To determine which of our three models best fits the data, we compare the predicted magnetic field strengths for each of the models to these observed values. 

\subsection{The toroidal morphology}

\label{ToroidalResult}
We investigate the Toroidal model using equation~\ref{HelicalBEq} and setting $B_1=0$ (i.e., a helix with zero pitch angle). By considering the ratios of each component of the  magnetic field to the total averaged magnetic field value, the free parameters  R and $B_0$ cancel out and do not contribute to the final results.

\begin{figure*}
\centering
\begin{tabular}{cc}
\includegraphics[scale=0.4, trim={0.25cm 0cm 1.5cm 1.5cm},clip]{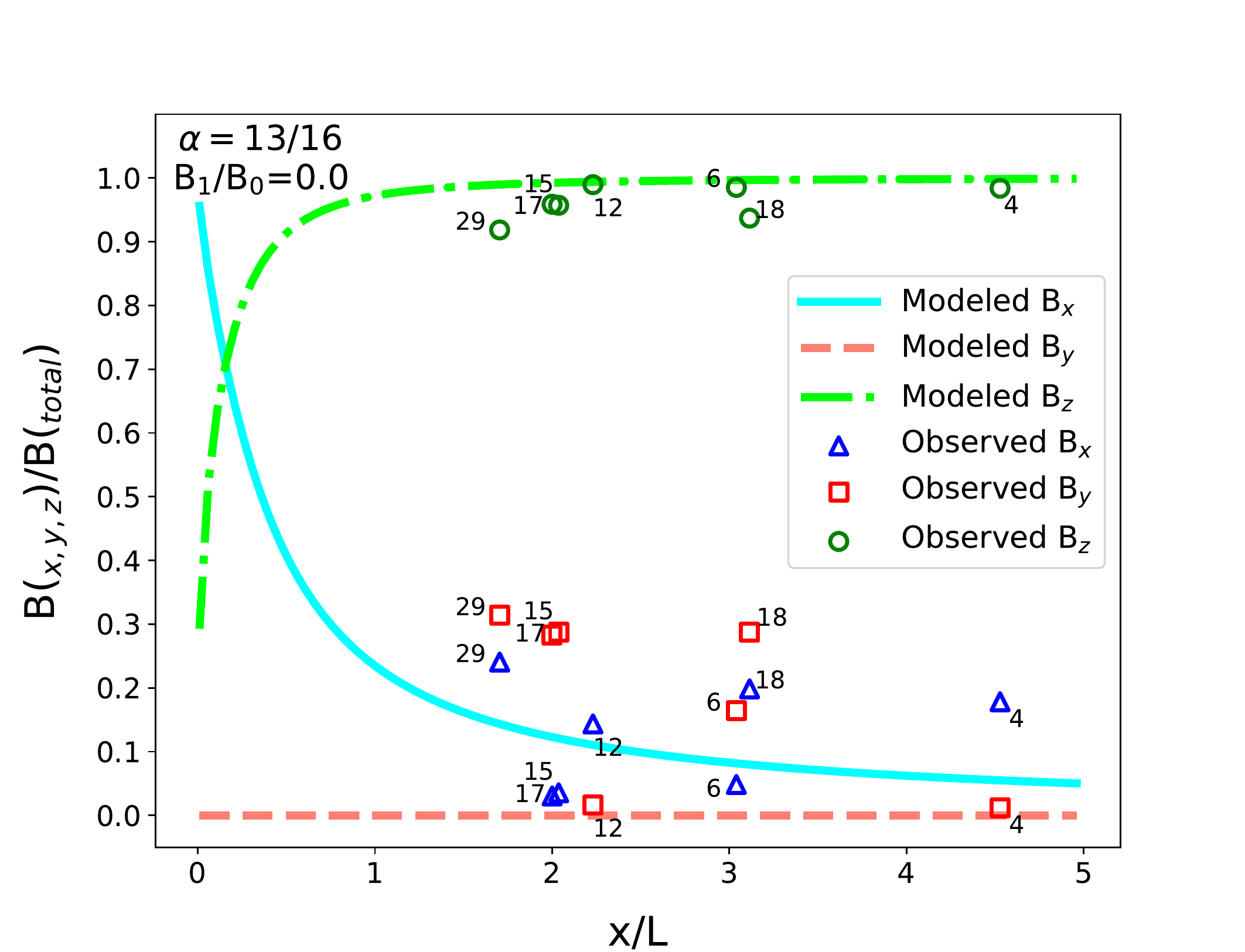}
\includegraphics[scale=0.4, trim={0.25cm 0cm 1.5cm 1.5cm},clip]{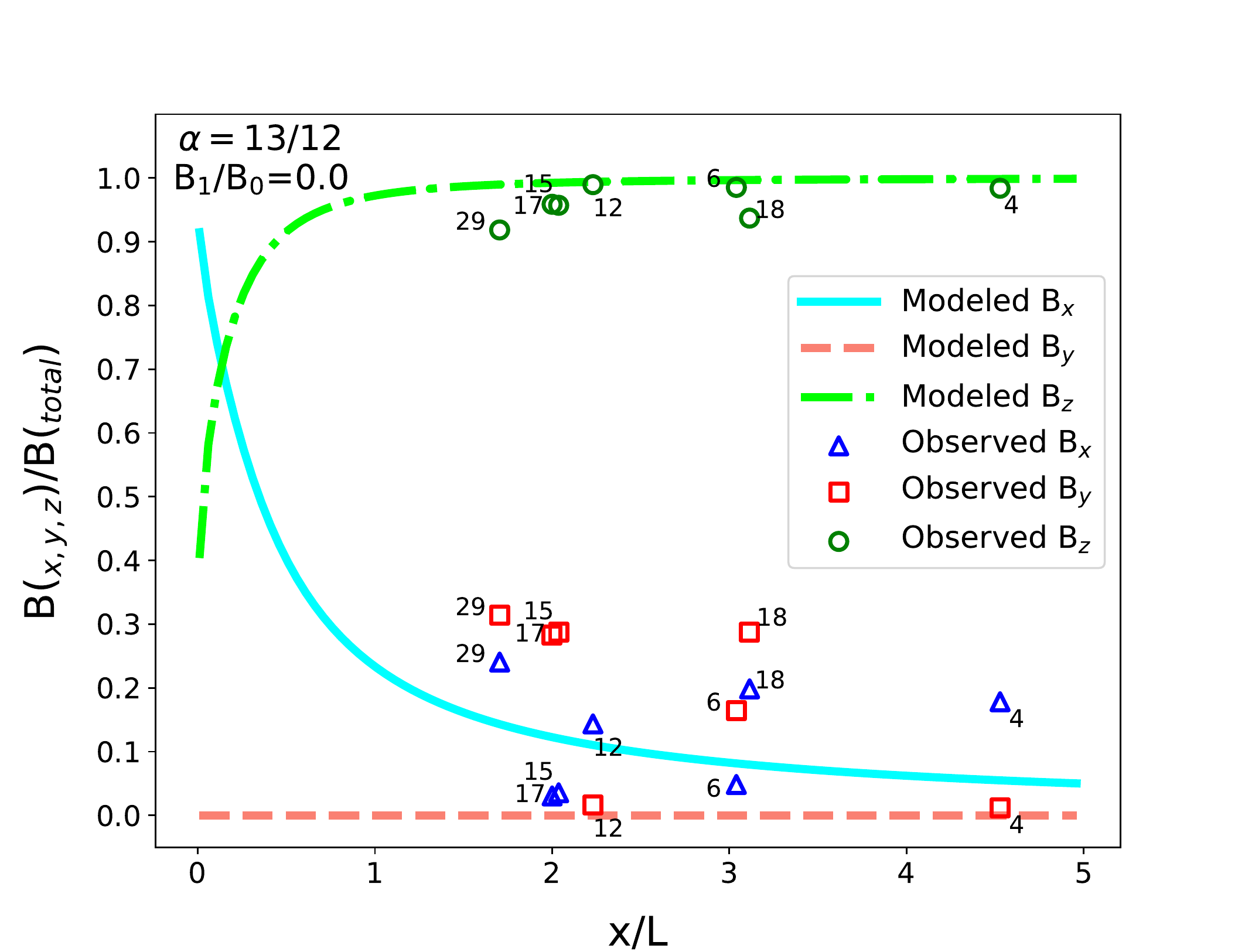}\\
\includegraphics[scale=0.4, trim={0.25cm 0cm 1.5cm 1.5cm},clip]{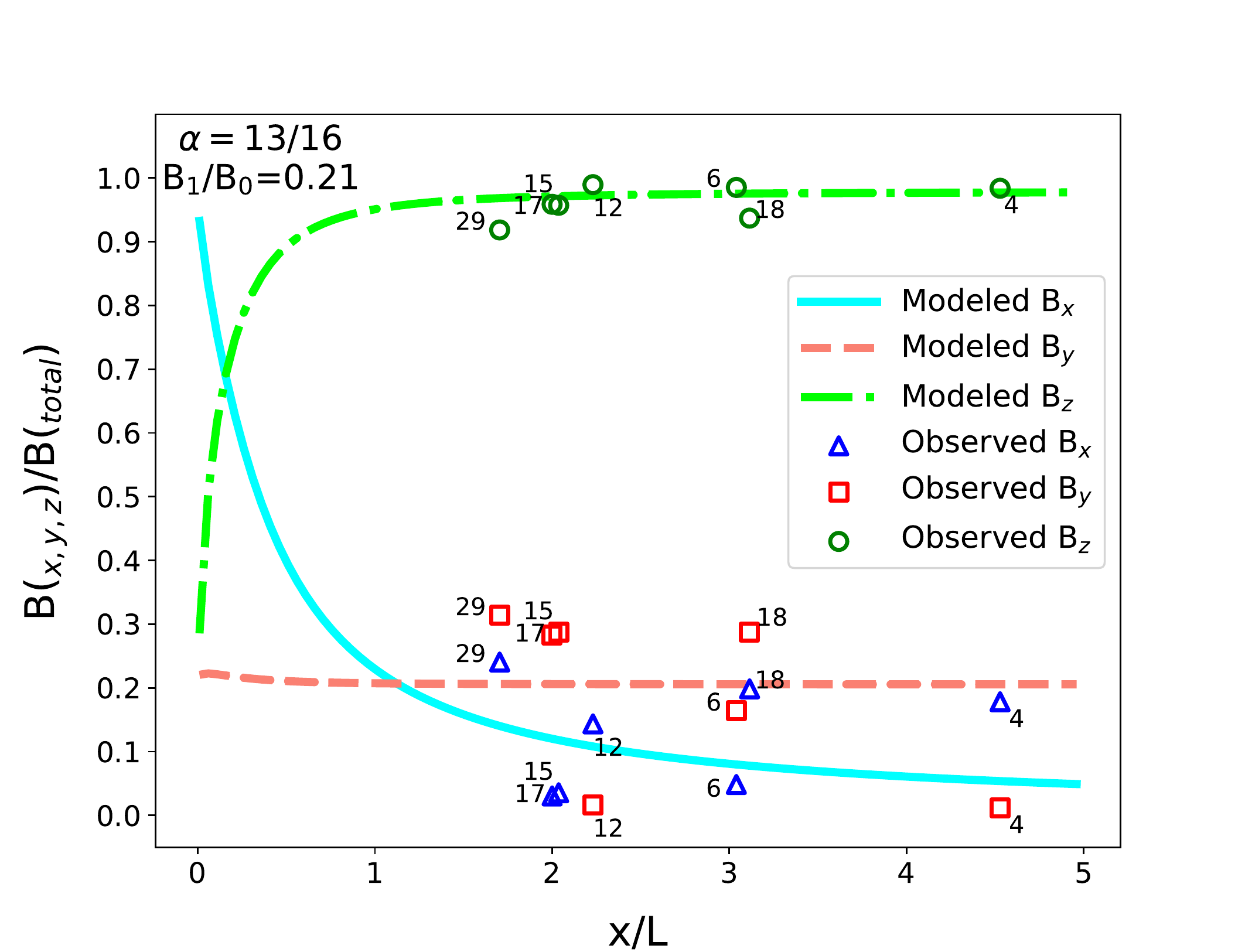}
\includegraphics[scale=0.4, trim={0.25cm 0cm 1.5cm 1.5cm},clip]{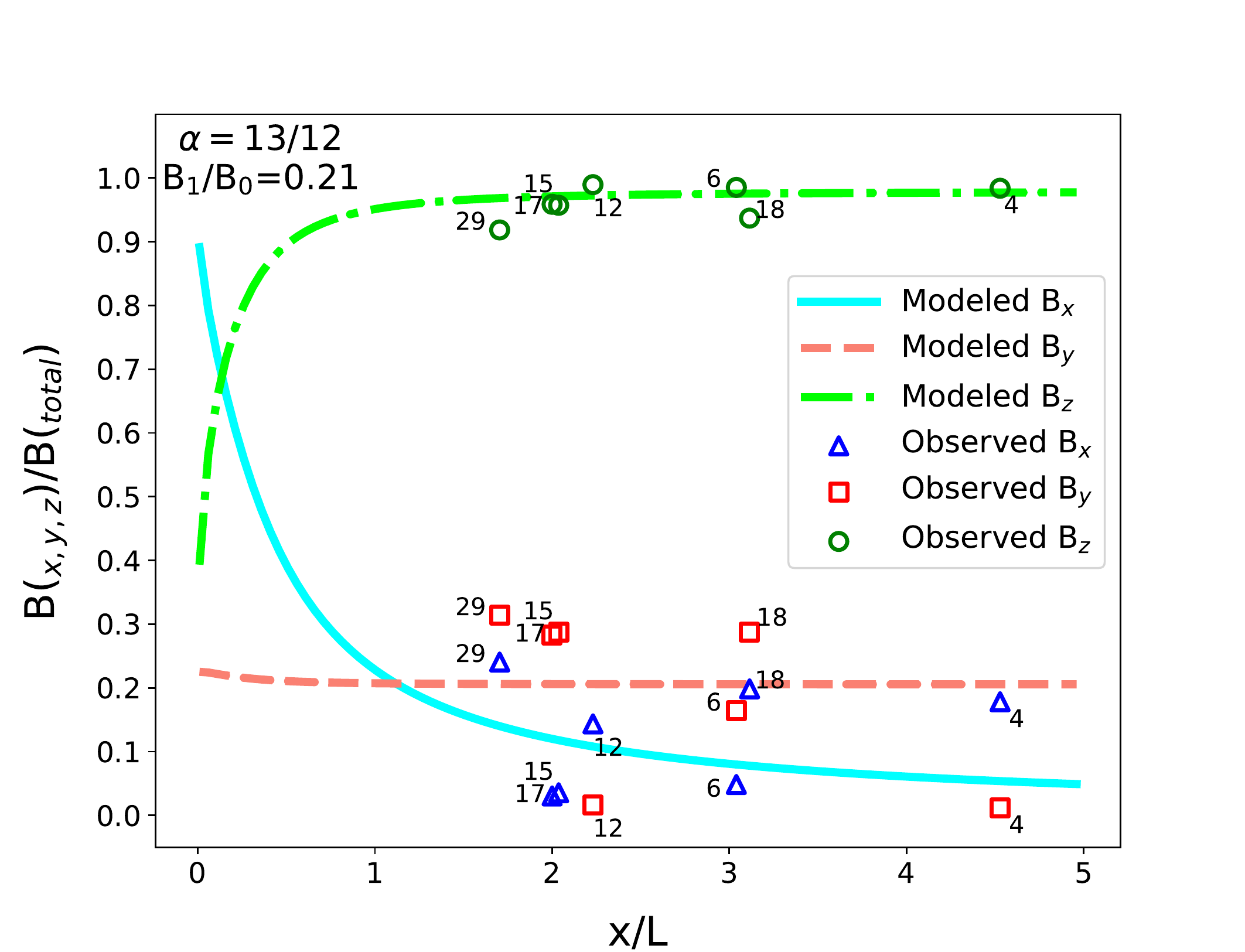}\\
\includegraphics[scale=0.4, trim={0.25cm 0cm 1.5cm 1.5cm},clip]{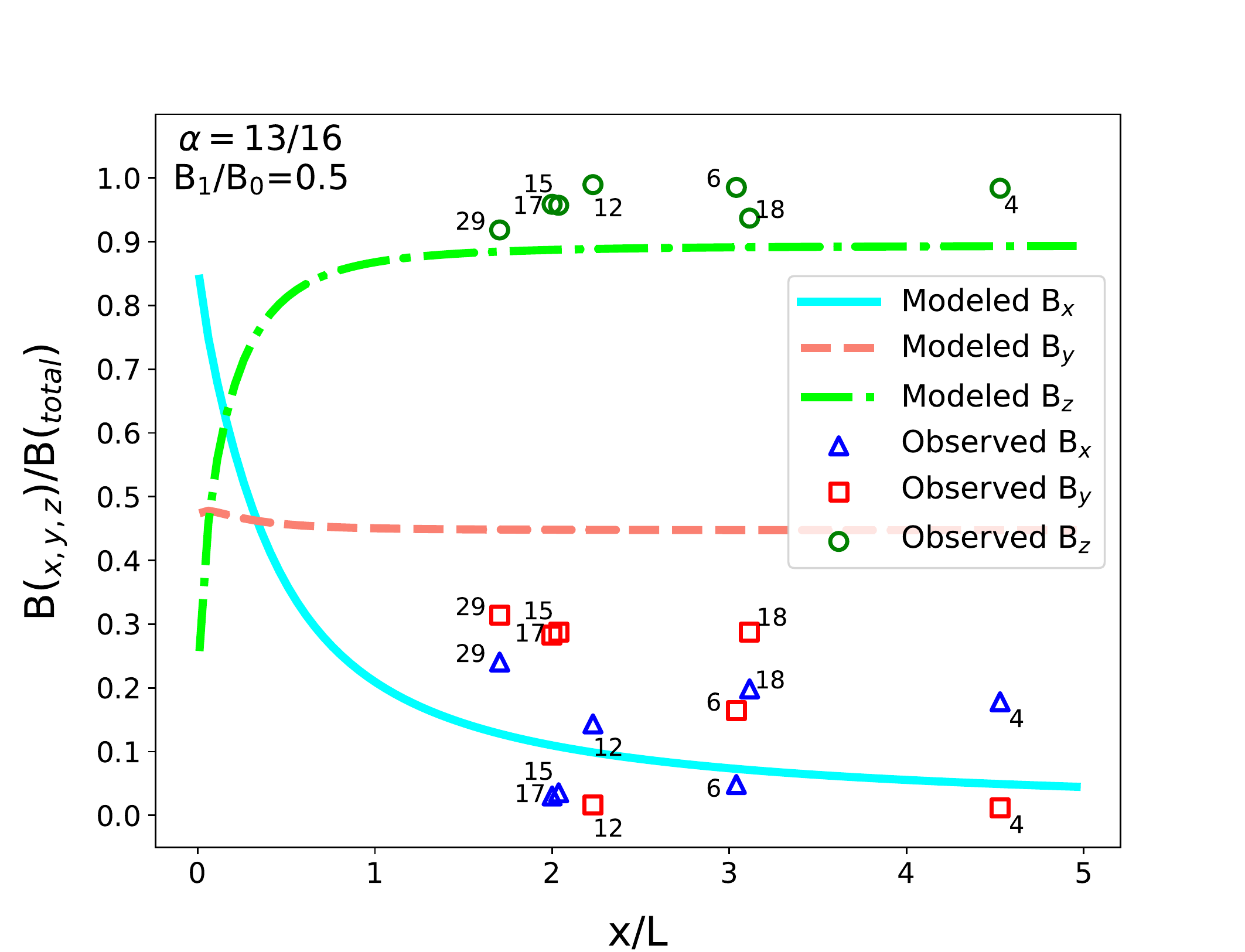}
\includegraphics[scale=0.4, trim={0.25cm 0cm 1.5cm 1.5cm},clip]{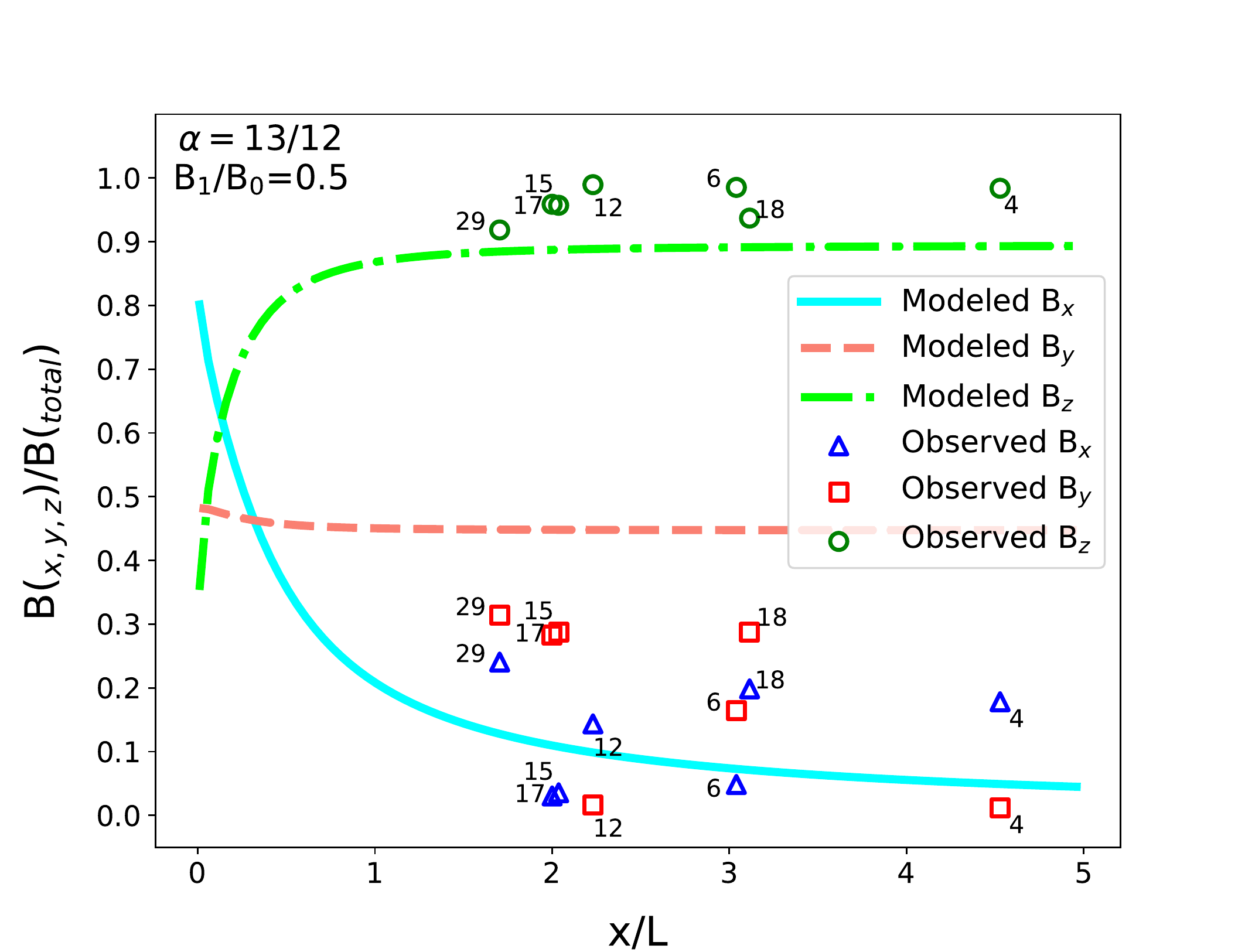}\\
\end{tabular}
\caption[Best fits for modeled toroidal and helical fields]{Best fits for modeled toroidal and helical fields. 
Colored lines represent the modeled x,y, and z components of the magnetic field. The colored symbols represent the x, y, and z  components of the observed magnetic field. 
On the x-axis, X is the perpendicular distance from the filament's long axis, and L is the LOS integration distance. The two $\alpha$ values are the parameters explained in Sec.~\ref{HelicalModel}.  The different $B_1$/$B_0$ ratios arise from equation~\ref{HelicalBEq} and set the level of helicity, with $B_1$/$B_0 = 0$ indicating a purely  toroidal field. 
} 
\label{ModeledBRatios}
\end{figure*}

The top row of Fig.~\ref{ModeledBRatios} shows the results of the toroidal model for the two different assumptions of the relationship between magnetic field strength and density: alpha = 13/16 (left) and 13/12 (right). The symbols represent the observed values, and the lines depict the model results as described in Sec.~\ref{ModelComparison}. The blue, red, and green colors illustrate the x, y, and z components, respectively. Both alpha values produce identical model results, which indicates that the decrease of magnetic field strength with distance does not play a noticeable role.

The top row of Fig.~\ref{ModeledBRatios} suggests that while the toroidal model seems to provide a good match to the z-component of the magnetic field, the modeled x- and y-components do not. The corresponding \KS\ value for each component are: $\chi ^2 _x = 0.109$, $\chi ^2 _y = 0.778$, $\chi ^2 _z = 0.020$, and $\chi ^2 _{tot} = 0.908$ for both  $\alpha$ = 13/16 and 13/12.

\subsection{The helical morphology}
\label{HelicalResult}
The pitch angle of a helix can be parameterised by the ratio of $B_1$/$B_0$ in equation~\ref{HelicalBEq}. Increasing this ratio increases the pitch angle.  After taking the LOS averages (see Sec.~\ref{ModelComparison}) $B_1$ and $B_0$ do not cancel out but instead remain as free parameters.
Thus, to study a helical model, we use equation~\ref{HelicalBEq} and vary the {\em ratio} of $B_1$/$B_0$,  from 0.0 to 1.49 in step sizes of 0.01.  For each value of $B_1$/$B_0$ we calculate the corresponding \KS\ value. The minimum \KS\ is calculated as the sum of the \KS\ for all the components (i.e., $\chi ^2 _{tot}$).  We also pay particular attention to the y-component, since it sets the toroidal model apart from the helical one.   
The lowest  \KS\ occurs for $B_1$/$B_0=0.21$ (both with $\alpha$ = 13/16 and 13/12), with  $\chi ^2 _x = 0.109$, $\chi ^2 _y = 0.193$, $\chi ^2 _z = 0.008$, $\chi ^2 _{tot} = 0.310$. The two bottom rows of Fig.~\ref{ModeledBRatios} represent the modeled helical morphology for $\alpha$ = 13/16 (left) and 13/12 (right). The middle, and the bottom rows show $B_1$/$B_0$ value of 0.21 (the best \KS\ fit result) and 0.5, respectively. 

\subsection{The bow morphology}
\label{BowAnalysis}

\begin{figure*}
\centering
\begin{tabular}{cc}
\includegraphics[scale=0.4, trim={0.25cm 0cm 1.5cm .5cm},clip]{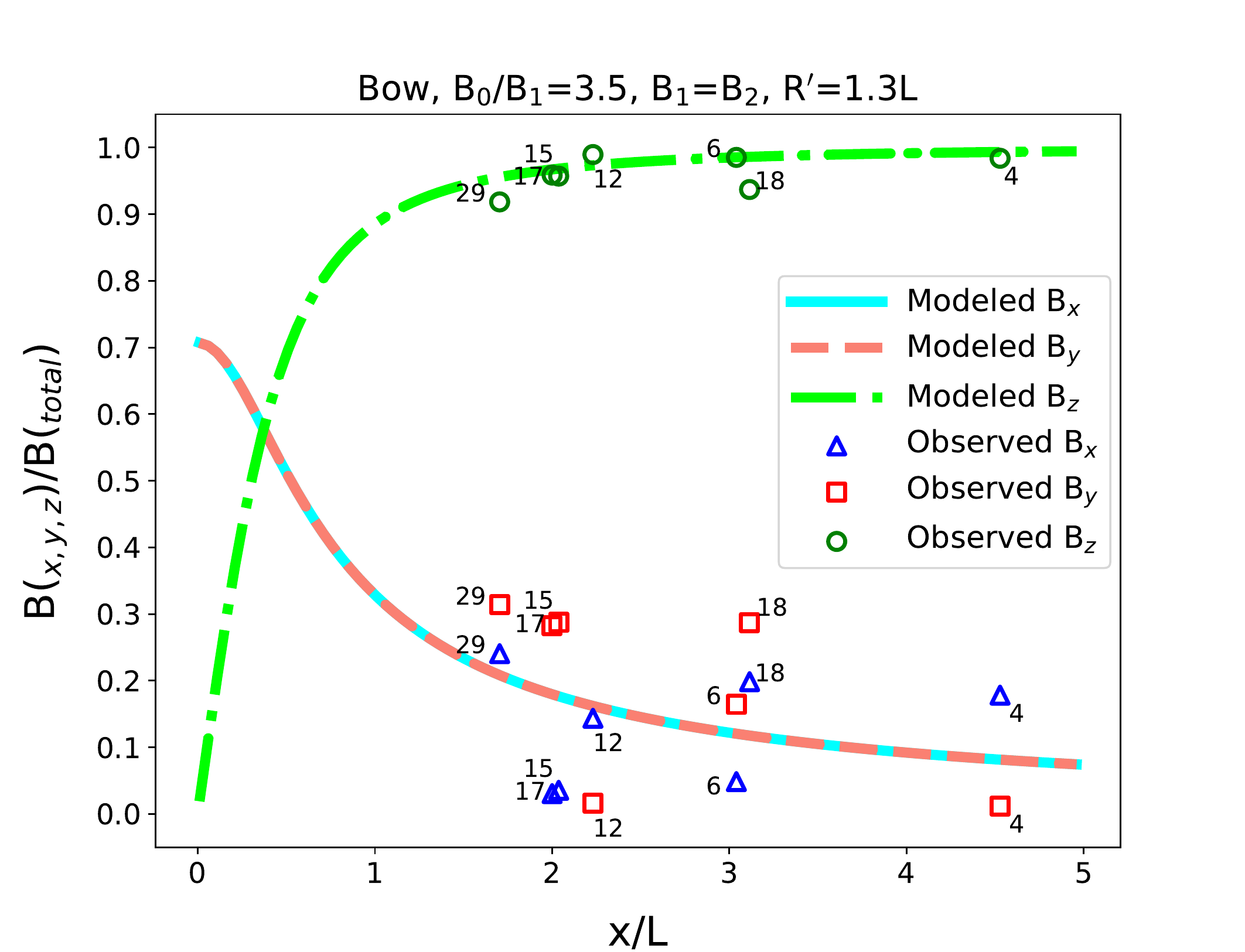}
\includegraphics[scale=0.4, trim={0.25cm 0cm 1.5cm .5cm},clip]{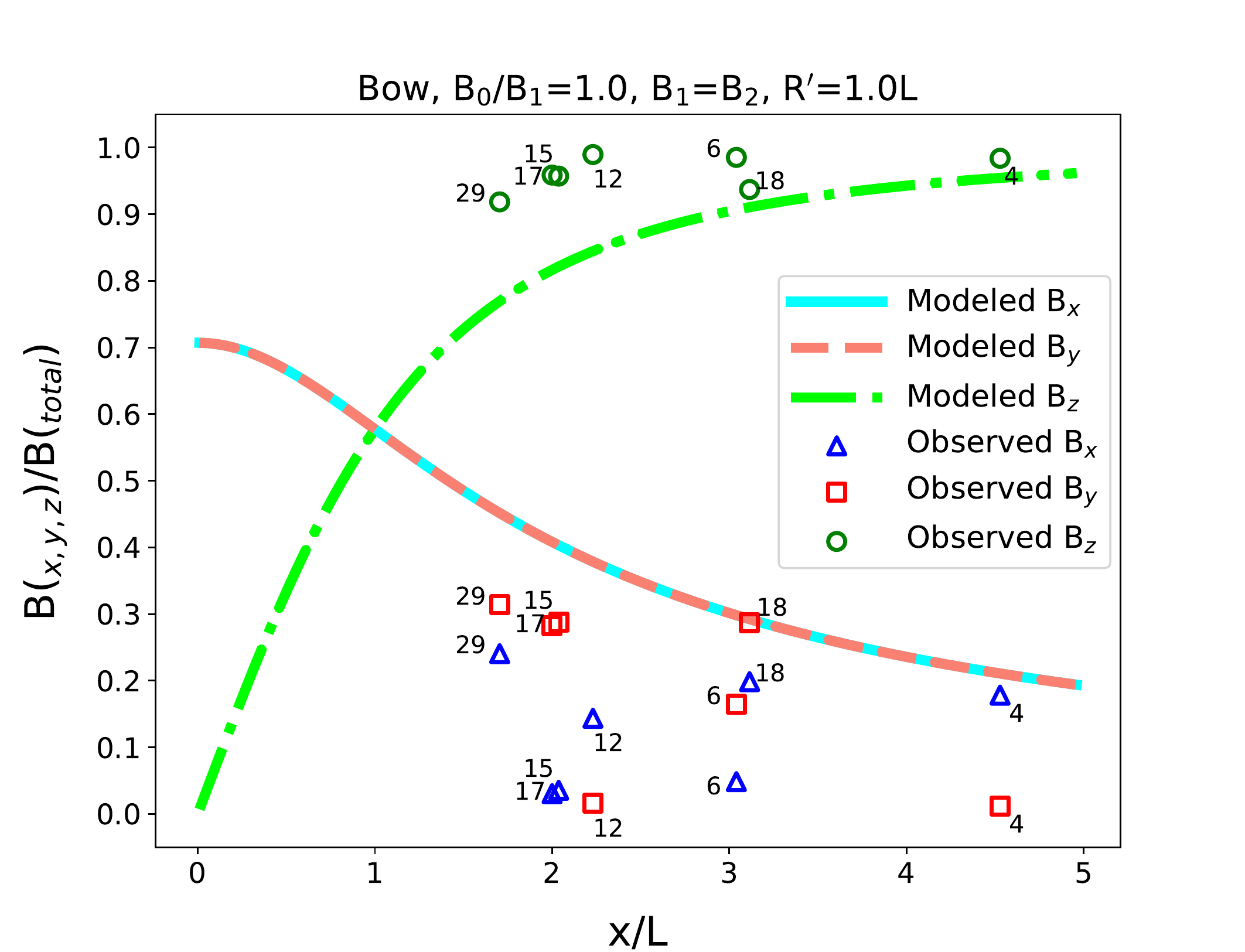}
\end{tabular}
\caption[Best fit tests of bow magnetic morphology.]{Bow magnetic morphology from equation~\ref{WrappedB}. The magnetic fields are averaged along the LOS.
Colored lines represent the modeled x,y, and z components of the magnetic field. The colored symbols represent the x, y, and z  components of the observed magnetic field. 
On the x-axis, X is the perpendicular distance from the filament's long axis, and L is the LOS integration distance. The parameters show different $B_1$/$B_0$ ratios from equation~\ref{WrappedB}. Left panel shows the best fit for equation~\ref{WrappedB}. }
\label{WrappedBRatios}
\end{figure*}

To model the bow magnetic morphology, we use equation~\ref{WrappedB} and find the ratio of each magnetic field component to the total magnetic field (as explained in Sec.~\ref{ModelComparison}).  Since the free parameters $B_1$, $B_0$, and R$^{\prime}$ do not cancel out, we explore different values of $B_0$/$B_1$ and R$^{\prime}$. To explore these free parameters, we first pick arbitrary values that result in a best visual fit with the data. For R$^{\prime}$, this initial value is equal to the integration distance, i.e., half of the perpendicular distance of point~17 from the filament axis. We then explore the influence of larger and smaller values of these free parameters in a quantitative manner. In addition, in  equation~\ref{WrappedB}, we assume $B_2$ = $B_1$, so that the x and y components of the magnetic field are equal, and then explore a range to find the lowest \KS\ results. Since the observed \blos\ values are higher than the Planck \bperp\ average strength, we start from $B_1$/$B_0=0.5$ and calculate the \KS\ for various R$^{\prime}$ values up to $B_1$/$B_0=15$, with step-size of 0.5. We take this range of parameters because we find that they fit the data better visually. In this range of $B_1$/$B_0$, we explore the R$^{\prime}$ values from 0.3 of the integration distance (0.3L) up to 3.2L, with step-size of 0.1. Fig.~\ref{WrappedBRatios} shows the results for this bow model for two sets of free parameters, the left panel represents a set of parameters that produce the smallest \KS\ result. The best-fit model is for $B_0$/$B_1=3.5$ and R$^{\prime}=1.3$L with $\chi ^2 _x= 0.158$, $\chi ^2 _y= 0.181$, $\chi ^2 _z = 0.007$, $\chi ^2 _{tot} = 0.346$.
 
To explore the bow model further, we consider equation~\ref{Reissl}, as described in Sec.~\ref{WaveModel}. For this purpose we modify the equations slightly to represent a more general form and to be compatible with our filament setup (i.e., to account for the reversal in the scales observed in \us\ and to make sure the units in the right and left hand sides of equation~\ref{Reissl} are the same [$\mu$G]).  We set U(x,z) in equation~\ref{ReissleInfo} as:
\begin{equation}
\begin{aligned}
& U(x,z) =  b_Rx(c_R-z)^2\exp(a_Rx^2),
\label{ReisslU}
\end{aligned}
\end{equation}
where a$_R$, b$_R$, and c$_R$ are free parameters. These parameters help the observed and modeled results to be close to each other visually.

\begin{figure*}
\centering
\begin{tabular}{cc}
\includegraphics[scale=0.4, trim={0.25cm 0cm 1.5cm 1.5cm},clip]{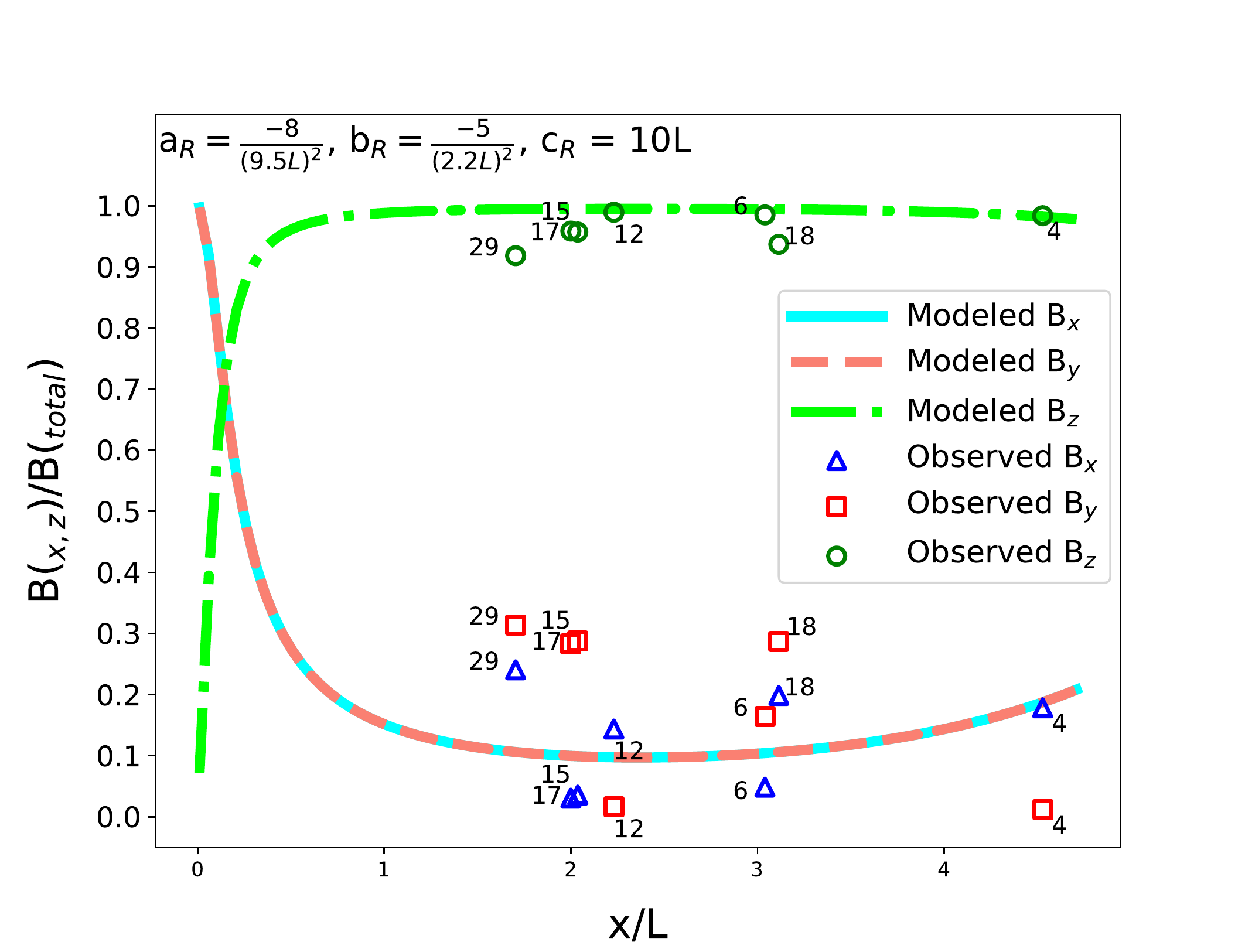}
\includegraphics[scale=0.4, trim={0.25cm 0cm 1.5cm 1.5cm},clip]{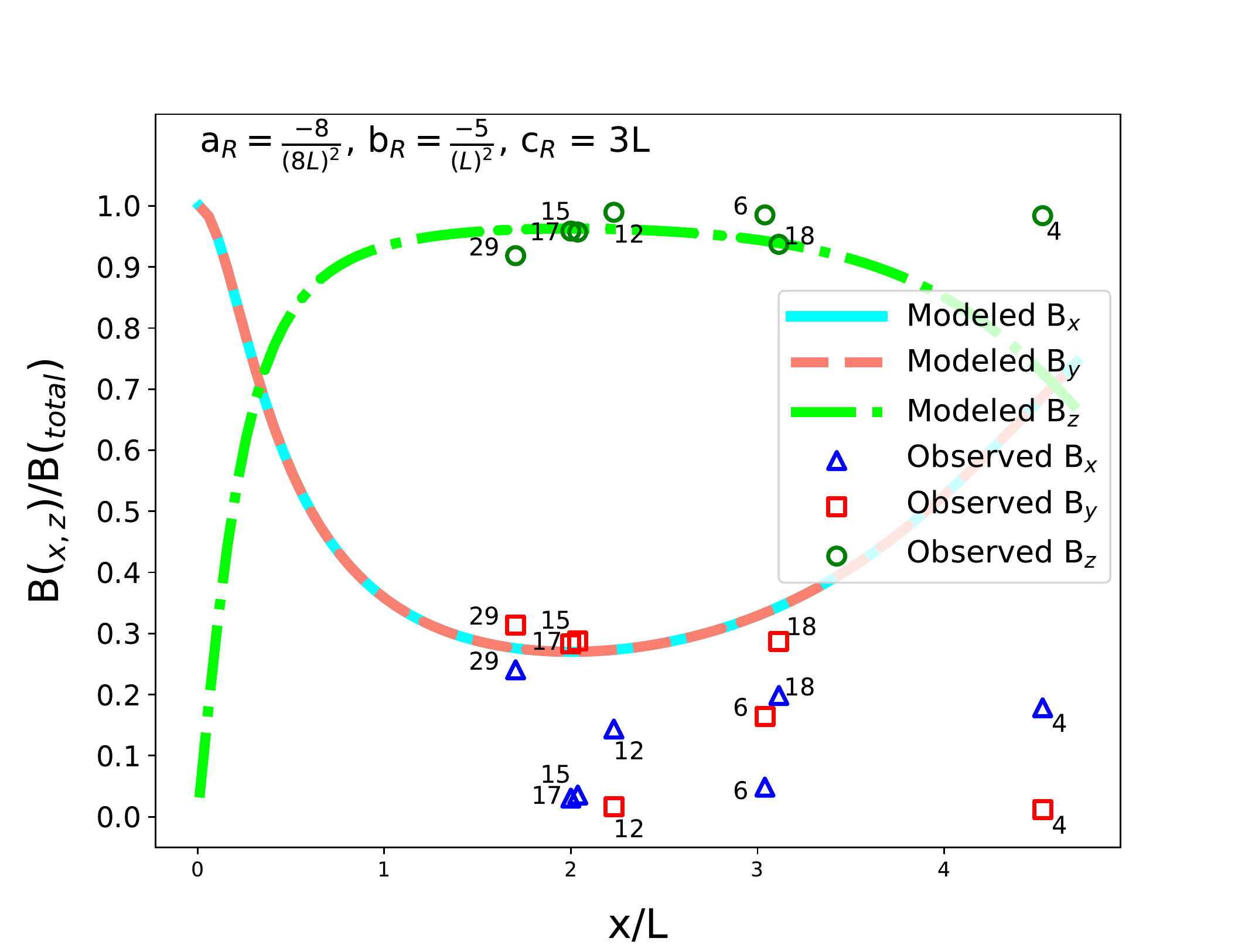}\\
\end{tabular}
\caption[Modeled bow magnetic morphology compared with data]{Bow magnetic morphology from equation~\ref{Reissl}. The magnetic fields are averaged along the LOS. 
Colored lines represent the modeled x,y, and z components of the magnetic field. The colored symbols represent the x, y, and z  components of the observed magnetic field. 
On the x-axis, X is the perpendicular distance from the filament's long axis, and L is the LOS integration distance. Left panel shows the best fit for equation~\ref{Reissl}. }
\label{ReisslBRatios}
\end{figure*}

To explore a range of free parameters, we set a$_R$, b$_R$, and c$_R$ in form of $\frac{-8}{(a\text{L})^2}$, $\frac{-5}{(b\text{L})^2}$, and c$_R=c$L, where a, b, and c are scalar values. In order to reduce the number of free parameters, we take the parameters as a function of L. These forms result in a unit of $\mu$G in equation~\ref{Reissl}. We subsequently explore different a, b, and c values where $ 5 \leq a \leq 17$ with step-sizes of 0.5, $ 0.2 \leq b \leq 3$ with step-size of 0.2, and $ 1 \leq c \leq 15$ with step-sizes of 1.
Fig.~\ref{ReisslBRatios} shows the results for
[a$_R=\frac{-8}{(9.5\text{L})^2}$, b$_R=\frac{-5}{(2.2\text{L})^2}$, and c$_R=10$L] and for [a$_R=\frac{-8}{(8\text{L})^2}$, b$_R=\frac{-5}{\text{L}^2}$, and c$_R=3$L]. We notice that different values of r$_{flat}$ do not generate any noticeable differences in final results, since we consider the ratios. 

Since the B$_y$ component is zero in this version of bow model,  $\chi _y ^2$ is constant and determined to be 0.778. This is lower than that of the Toroidal model, but higher than those of both the helical and bow models from equation~\ref{WrappedB}. We find that a few sets of parameters with $a= 9.5$ result in the lowest \KS with c in a range of 9 to 13 and b in range of 2.0 to 2.8. For example a = 9.5, b = 2.2, and c = 10 (i.e., a$_R=\frac{-8}{(9.5\text{L})^2}$, b$_R=\frac{-5}{(2.2\text{L})^2}$, and c$_R=10$L) or a = 9.5, b = 2.4, and c = 11 produce \KS\ result of $\chi ^2 _x = 0.073$, $\chi ^2 _y = 0.778$, $\chi ^2 _z = 0.021$, $\chi ^2 _{tot} = 0.872$. Given the small step-sizes having a range for the smallest \KS\ is expected. For this bow model, we pick the set of parameter that when we consider more significant digits of \KS , they result in the smallest \KS . That is with a$_R=\frac{-8}{(9.5\text{L})^2}$, b$_R=\frac{-5}{(2.2\text{L})^2}$, and c$_R=10$L.

Finally, we should note that in equations~\ref{Reissl} and~\ref{ReissleInfo}, the density is that of a Plummer-like profile to relate magnetic strength to r in $B_0(x,z)$. An alternative approach is to use the earlier proposed relation between the magnetic strength and r (see Sec.~\ref{HelicalModel}) of
\begin{equation}
\begin{aligned}
& B_0(x,z) =  \frac{B_0}{(x^2+z^2)^{\frac{-\alpha}{2}}}.
\label{dropOff}
\end{aligned}
\end{equation}
However, even with this density profile, equation~\ref{Reissl} for this bow model is not divergence-free. We find that using the two different density profile does not provide noticeably different results.

Finally, we note that these results are based on magnetic field values that have large uncertainties and possibly have systematic biases. To  alleviate the dependency of our interpretation of the 3D magnetic field morphology on sole values of \bperp\ and \blos\, we carry out a Monte-Carlo analysis in Sec.~\ref{discussion}.

\section{Discussion}
\label{discussion}

In this section we explore the results further and  propose the most likely magnetic field morphology  for Orion-A. We then consider effects that may have biased our results.

\subsection{A Monte-Carlo analysis} 
\label{monteCarlo}

PXXXV provide a magnetic field value (\bperp)  that is averaged over the entire Orion-A region, whereas \us\ provide the \blos\ field values for very specific points with observed RMs. 
Thus, applying this single value of \bperp\ to every RM position comes with significant uncertainties. In addition, the values of \blos\ from \us\ come with their own, inherently large, uncertainties. To account for these uncertainties, we perform a Monte-Carlo error analysis to compare the models with the observed magnetic field components. In this approach, we randomly alter the \blos\ and \bperp\ strengths of each single data point separately within their uncertainty ranges and, for each iteration, compare the models with the data. More specifically, we
\begin{enumerate}
\item Randomly changed \blos\ for each data point within its uncertainty range (from \us).
\item Randomly changed the strength of Planck magnetic field (\bperp) for each data point separately within its uncertainty range ($\pm 90\,\mu$G). 
\item Calculated the ``new'' $B_x$, $B_y$, and $B_z$ values, using the ``new'' \bperp\ and \blos\ values for each data point.
\item Found the differences ($\varsigma$) between the ``new''  $B_x$, $B_y$, and $B_z$ values and the best toroidal/helical/bow models found in Sec.\ref{results} using the equations:
\begin{equation}
\begin{aligned}
& \varsigma _x  =  \sum_{i=1}^{N} (|B_{x, \text{ modeled}}| - |B_{x,\text{ observed}}|)_i^2,\\
& \varsigma _y  =  \sum_{i=1}^{N} (|B_{y, \text{ modeled}}| - |B_{y,\text{ observed}}|)_i^2,\\
& \varsigma _z  =  \sum_{i=1}^{N} (|B_{z, \text{ modeled}}| - |B_{z,\text{ observed}}|)_i^2,\\
& \varsigma _{\text{total}}  = \varsigma_x  + \varsigma_y + \varsigma_z,
\end{aligned}
\label{MCDiffs}
\end{equation}
where N is the number of data points (N = 7).
\item Found which of our best toroidal/helical/bow models  has the lowest $\varsigma_{\text{total}}$ value when compared against the ``new'' (i.e., randomly altered) data. 
\end{enumerate}
This Monte-Carlo analysis was done 50,000 times.  Our bow model had the smallest difference (i.e., the bow model best fits the data) 34,177 times.  The helical model had the smallest difference 15,823 times. In no cases did the toroidal model have the smallest difference.

\subsection{Systematic biases}

Since \bperp\ and \blos\ are derived in completely different ways (using Faraday rotation for \blos\ and dust polarization for \bperp), we probe the effects of potential systematic biases between the results of the two methods. Additionally, we should note that our values of \blos\ are generally larger than the \bperp\ value and the \bperp\ value is an average for the region. Therefore, our \bperp\ could be systematically lower than \blos. To investigate potential systematic biases, we explored a range of factors multiplied by our \blos\ and \bperp\ separately, and repeated a process similar to that presented in Sec.~\ref{monteCarlo}. In more details, we
\begin{enumerate}
\item Multiplied all of our \blos\ values by a randomly generated factor between 0.5 and 5 (same factor for all of the \blos\ values).
\item Multiplied the \bperp\ value by a randomly generated factor between 1 and 5. 
\item Calculated the ``new'' $B_x$, $B_y$, and $B_z$ values, using the ``new'' \bperp\ and \blos\ values for each data point.
\item Using equation~\ref{MCDiffs}, found the differences ($\varsigma$) between the ``new''  $B_x$, $B_y$, and $B_z$ values and the best toroidal/helical/bow models found in Sec.\ref{results}.
\item Found which of our best toroidal/helical/bow models  has the lowest $\varsigma_{\text{total}}$ value when compared against the ``new'' data. 
\end{enumerate}
This analysis was done 50,000 times.  Our bow model had the smallest difference (i.e., the bow model best fits the data) for 25002 times. The bow shape in form of equation~\ref{ReisslU} and the helical model had the smallest difference for 5195 and 19803 times, respectively. Therefore, considering this range of systematic biases, the bow model seems to be a better fit. 

\subsection{Selecting the best magnetic field morphology}

To compare the toroidal and helical models we only need to compare the \KS\ values of the two models with each other. This is because these two models are a single model with different parameters (one with $B_1$/$B_0=0$ and the other with $B_1$/$B_0=0.21$). Therefore, since $\chi ^2 _{tot} = $ 0.310 for the helical model and  0.908 for the toroidal model, we suggest a helical morphology fits the data better.  Furthermore, since it is the y-component that sets the toroidal and helical models apart, visually comparing the y-component  in the top panel of Fig.~\ref{ModeledBRatios} with the middle-left panel  illustrates that a helical model (with $B_1$/$B_0=0.21$) is a better fit to the data.  This is borne out in the mathematical analysis which shows that  $\chi ^2 _y = $ 0.193 for the helical model and 0.778 for the toroidal model.

To compare the bow morphology to the helical/toroidal one, we first compare the left panel of Fig.~\ref{WrappedBRatios} and the left panel of Fig.~\ref{ReisslBRatios} with the helical and toroidal models in Fig.~\ref{ModeledBRatios}. From these figures  it is easy to see that the bow models fit the data better than a toroidal model.  However, it is difficult to see if either of the bow models fit the data better than the helical model. Thus, a more robust comparison is required.

Since the bow and toroidal/helical morphologies are represented by two very different models, one cannot simply compare their \KS\ values. Thus, to compare the bow morphology to the toroidal/helical one, we compare their probability values (i.e., p-values). P-values allow one to compare different models by providing the probability for a statistically relevant model. The probability value provides the likelihood that a hypotheses (toroidal, helical, or bow) is true \citep{DataAnalysis}. In statistical analysis,  p-values are normally used to reject a null hypothesis. If we assume that they represent a Gaussian distribution, it is a measure to understand whether the result from the hypothesis is closer to the peak of the distribution or its tail. A model/hypothesis can be ruled out if the p-value is less than  a threshold, typically 0.05.

To find the p-values, we need the \KS\ values and the degrees of freedom (DOF) of the model. For the helical model we take one degree of freedom (i.e., $B_1$/$B_0$) for the first bow model we take two degrees of freedom  (i.e., $B_0$/$B_1$, and R$^{\prime}$) for the second bow model we take three degrees of freedom (i.e., a$_R$, b$_R$, and c$_R$). 
Our best fit bow model ($\chi ^2 _{tot}$ =0.346  with two DOFs) has a p-value of 0.84.  Our best helical model ($\chi ^2 _{tot}  = 0.310$ with one DOF) has a p-value of 0.58.
Based on these p-values none of the models can be ruled out. We should note that the \cite{Reissletal2018} bow model ($\chi ^2 _{tot}$ = 0.872 with three DOF) has a  p-value  of  0.83 - similar to that of our bow model.

Therefore, it is likely that the bow morphology is more probable than both the  helical and toroidal. Additionally, the results of the Monte-Carlo analysis and our systematic bias investigation lend support to our suggestion that the bow model is, overall, the most likely fit to the observed magnetic fields  in the Orion-A filament.

\begin{figure}
\centering
\hspace{-0.5cm}
\includegraphics[scale=0.9, trim={0.26cm 0cm 0cm 0cm},clip]{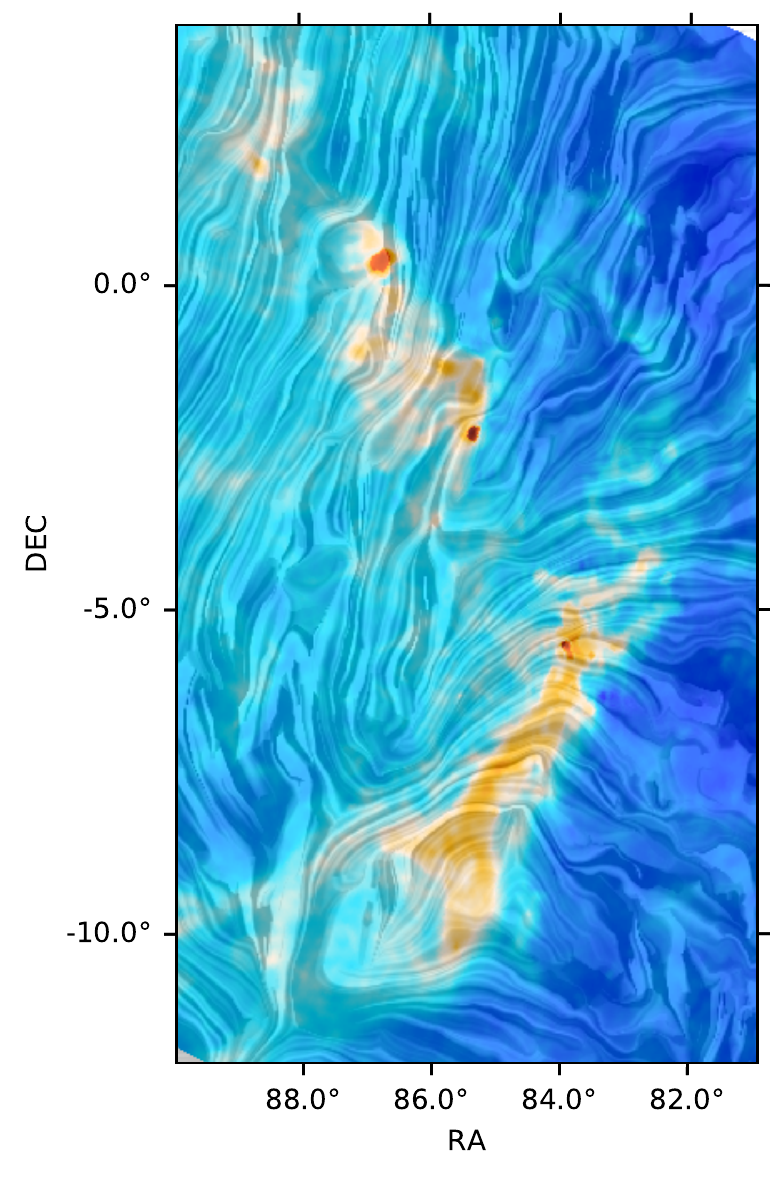}
\caption[Plane-of-sky component of magnetic field in Orion as obtained by Planck observations]{Column density and magnetic field toward the Orion region. The colors correspond to the dust opacity derived from the Planck observations.
The drapery pattern 
corresponds to the \emph{orientation} of the magnetic field projected on the plane of the sky and integrated along the line of sight, as inferred from the Planck 353-GHz observations. The yellow color between declination of about $-5^\circ$ and $-10^\circ$ show the Orion-A filamentary structure as illustrated in Fig.~\ref{OrionBlos}. The \bperp\ lines on one side of this filament are mostly perpendicular to the filament (representing a small pitch angle for a helical morphology) while, on the other side, they are more parallel to the filament. 
} 
\label{OrionPlanck}
\end{figure}

Furthermore, as shown by the Planck results in Fig.~\ref{OrionPlanck}, the \bperp\ lines on one side of the Orion-A filament are mostly perpendicular to the filament (representing a small pitch angle for a helical morphology) while, on the other side, they are more parallel to the filament.  Therefore, if we assume a helical magnetic field morphology for the Orion-A region, we would have to consider completely different pitch angles for the two sides of the filament which is not sensible.  Therefore, this simple visual analysis of the \bperp\ orientation also seems to suggest that a bow morphology is a more natural fit to the data.

\subsection{Tilted models for Orion-A}

Finally, we briefly explore the effects of 3D geometry on our results. Recent observations by \cite{Grossschedletal2018} show that the Orion-A filamentary structure is inclined along the LOS with respect to the observer. However, they show that different parts of this region have different inclination angles. The part of the Orion-A region that contains Orion KL has a close-to-zero tilt. In particular, the region that contains L1641-N in Orion-A, which is very close to 6 of the 7 data points taken in  our  analysis, has zero inclination in Fig.~3 of \cite{Grossschedletal2018}. Due to this zero inclination at the region where our data points are accumulated, we suggest that we do not need to consider inclination of the Orion-A region along the line-of-sight  for this study. 

Consequently, the models of equations~\ref{HelicalBEq}, \ref{WrappedB}, and \ref{Reissl} represent a filament that is assumed to be lying in the plane of the sky without being tilted away from, or towards, the observer. With future data sets, where the data points are accumulated all along the Orion-A region, considering a more complex 3D geometry incorporating the study of \cite{Grossschedletal2018}, is required. In that case, if one considers a tilted filament, the equations representing the models will not change since the frame of reference is adjusted to the filament as illustrated in Fig.~\ref{Cylindre}. However, when taking the average of each magnetic field component along the line-of-sight, one has to adjust $dz$ in equation~\ref{average}, such that this $dz$ is the line-of-sight path length from the observer's point of view and not in the filament's reference frame.

\section{Conclusions}
\label{conclusion}

We present 3D magnetic field modelling around the Orion-A molecular cloud using existing dust polarization observations, which provide the plane-of-sky magnetic field, combined with the line-of-sight magnetic fields obtained by the method presented in \cite{Tahanietal2018}.
We first construct models describing a toroidal, helical, and a bow-shape magnetic field morphology. While all  three morphologies can explain the line-of-sight magnetic field reversal around Orion-A as observed by~\cite{Tahanietal2018}, we  suggest that the bow morphology is the most natural magnetic field morphology for the Orion-A region. 

This conclusion is based on the following arguments:
\begin{enumerate}
\item The bow model has the highest p-values.
\item Our bow model seems to be the most consistent model with the observational data and their uncertainty values in our Monte-Carlo analysis approach, as well as with our investigation of systematic biases.
\item Different \bperp\  angles on the eastern side of the filament compared to the western side 
suggests that the data cannot be properly modeled by a single helix with the same pitch angle on both sides of the filament. 

\end{enumerate}

The analysis in this paper utilises the best rotation measure data currently available in the entire Orion-A region.  As such, it is limited to the number of available RM sources in the \cite{Tayloretal2009} catalog, as well as the sensitivity of those observations.  This analysis is also limited by the fact that the Planck maps provide only one value for \bperp\ averaged across the entire region. Our results stress the need for more RM observations and higher-angular resolution dust polarization maps to further reconstruct the 3D magnetic field morphology. 

Future RM catalogs from new generation surveys such as POSSUM and VLASS will allow for the detection of a  higher number of \blos\ sources and with better sensitivity. Thus, these new catalogs will provide better-sampled maps of the large-scale  \blos\ magnetic field with smaller errors.  Combined with new and improved dust polarization maps (from instruments like APEX, ALMA, BLASTPOL2), future applications of the analyses presented in this work will allow us to map and model the 3D magnetic field geometry with better accuracy.

\begin{acknowledgements}
We thank the anonymous referee for his/her insightful comments that helped improve this paper. We have used LATEX, Python and its associated libraries, Pycharm, and SAO Image DS9 for this work. We have used Magnetar package
\footnote{github.com/solerjuan/magnetar} for line integral convolution
(LIC) magnetic field visualization in Fig.~\ref{OrionPlanck}. MT thanks Anahita Alavi for a discussion on statistical approaches. JDS acknowledges funding from the European Research Council under the Horizon 2020 Framework Program via the ERC Consolidator Grant CSF-648505. JK has received funding from the European Union’s Horizon 2020 research and innovation program under grant agreement No 639459 (PROMISE). We acknowledge the Planck collaboration for the publicly-available data through the Planck Legacy Archive.

\end{acknowledgements}

\bibliographystyle{aa} 
\bibliography{bibt}

\end{document}